\documentclass[%
 reprint,%
 amssymb, amsmath,%
 aip,cha,%
]{revtex4-1}

\usepackage{docs}%
\usepackage{bm}%
\usepackage{bookmark}

\expandafter\ifx\csname package@font\endcsname\relax\else
 \expandafter\expandafter
 \expandafter\usepackage
 \expandafter\expandafter
 \expandafter{\csname package@font\endcsname}%
\fi
\usepackage{arydshln}
\usepackage{epsfig}
\usepackage{url}
\usepackage{amssymb, amsmath, amsthm} 
\usepackage{epsfig,graphicx} 
\usepackage{subcaption}
\usepackage{xspace}
\usepackage{arydshln}
\usepackage{color}
\usepackage{verbatim}
\usepackage{pifont} 
\usepackage{color}
\usepackage{listings}
\begin{document}

\title{Prediction in Projection}

\author{Joshua Garland}%
\email{joshua.garland@colorado.edu}
\affiliation{University of Colorado\\Department of Computer Science\\ Boulder, Colorado 80303, USA}%
\author{Elizabeth Bradley}%
\email{lizb@colorado.edu}
\affiliation{University of Colorado\\Department of Computer Science, Boulder, Colorado 80303, USA \\ and the Santa Fe Insitute, Santa Fe, New Mexico}%
\date{}%
\revised{}%

\maketitle

\newcommand{\cmark}{\ding{51}}
\newcommand{\xmark}{\ding{55}}
\newcommand{\alert}[1]{{\color{red}#1}}
\newcommand{\espace}[2]{\ensuremath{\mathbb{E}[#1,#2]}}

\newcommand{\cmt}[1]{\textcolor{magenta}{#1}}

\newtheorem*{mydef}{Definition}
\newtheorem{thm}{Theorem}
\newtheorem*{remark}{Remark}
\newtheorem{prop}[thm]{Proposition}
\newtheorem{example}[thm]{Example}


\newcommand{\gcc}{{\tt 403.gcc}\xspace}
\newcommand{\sphinx}{{\tt 482.sphinx}\xspace}
\newcommand{\eig}{{\tt dgeev}\xspace}

\newcommand{\naive}{na\"ive\xspace}
\newcommand{\col}{{\tt col\_major}\xspace}

\newcommand{\citeNTau}{\cite{fraser-swinney,kantz97,Buzug92Comp,liebert-wavering,Buzugfilldeform,Liebert89,rosenstein94}\xspace}

\newcommand{\citeNM}{\cite{liebert-wavering,Cao97Embed,Kugi96,Buzugfilldeform,KBA92,Hegger:1999yq,kantz97}\xspace}

\newcommand{\citeNEPARAMS}{\cite{fraser-swinney,kantz97,Buzug92Comp,liebert-wavering,Buzugfilldeform,Liebert89,rosenstein94,Cao97Embed,Kugi96,KBA92,Hegger:1999yq}}

\newcommand{\citeDCEFORECASTING}{\cite{weigend-book,casdagli-eubank92,Smith199250,pikovsky86-sov,sugihara90,lorenz-analogues}}


\newcommand{\fnnLMA}{{\tt fnn-LMA}\xspace}
\newcommand{\roLMA}{{\tt ro-LMA}\xspace}
\newcommand{\ipc}{{\tt IPC}\xspace}

\section*{Abstract}

Prediction models that capture and use the structure of state-space
dynamics can be very effective.  In practice, however, one rarely has
access to full information about that structure, and accurate
reconstruction of the dynamics from scalar time-series data---e.g.,
via delay-coordinate embedding---can be a real challenge.  In this
paper, we show that forecast models that employ \emph{incomplete}
embeddings of the dynamics can produce surprisingly accurate
predictions of the state of a dynamical system.  In particular, we
demonstrate the effectiveness of a simple near-neighbor forecast
technique that works with a two-dimensional embedding.  Even though
correctness of the topology is not guaranteed for incomplete
reconstructions like this, the dynamical structure that they capture
allows for accurate predictions---in many cases, even more accurate
than predictions generated using a full embedding.  This could be very
useful in the context of real-time forecasting, where the human effort
required to produce a correct delay-coordinate embedding is prohibitive.


\section*{Lead Paragraph}

Prediction models constructed from state-space dynamics have a long
and rich history, dating back to roulette and beyond.  A major
stumbling block in the application of these models in real-world
situations is the need to reconstruct the dynamics from scalar
time-series data---e.g., via delay-coordinate embedding.  This
procedure, which is the topic of a large and active body of
literature, involves estimation of two free parameters: the dimension
$m$ of the reconstruction space and the delay, $\tau$, between the
observations that make up the coordinates in that space.  Estimating
good values for these parameters is not trivial; it requires the
proper mathematics, attention to the data requirements, computational
effort, and expert interpretation of the results of the calculations.
This is a major challenge if one is interested in real-time
forecasting, especially when the systems involved operate on fast time
scales.  In this paper, we show that the full effort of
delay-coordinate embedding is not always necessary when one is
building forecast models, and can indeed be overkill.  Using synthetic
time-series data generated from the Lorenz-96 atmospheric model and
real data from a computer performance experiment, we demonstrate that
a two-dimensional embedding
of scalar time-series data from a dynamical system gives simple
forecast methods enough traction to generate accurate predictions of
the future course of those dynamics---sometimes even more accurate
than predictions created using the full embedding.  Since incomplete
embeddings do not preserve the topology of the full dynamics, this is
interesting from a mathematical standpoint.  It is also potentially
useful in practice.  This reduced-order forecasting strategy involves
only one free parameter ($\tau$), good values for which, we believe,
can be estimated `on the fly' using information-theoretic and/or
machine-learning algorithms.  As such, it sidesteps much of the
complexity of the embedding process---perhaps most importantly, the
need for expert human interpretation---and thus could enable
automated, real-time dynamics-based forecasting in practical
applications.

\section{Introduction}\label{sec:intro}

Complicated nonlinear dynamics are ubiquitous in natural and
engineered systems.  Methods that capture and use the state-space
structure of a dynamical system are a proven strategy for forecasting
the behavior of systems like this, but the task is not
straightforward.  The governing equations and the state variables are
rarely known; rather, one has a single (or perhaps a few) series of
scalar measurements that can be observed from the system.  It can be a
challenge to model the full dynamics from data like this, especially
in the case of \emph{forecast} models, which are only really useful if
they can be constructed and applied on faster time scales than those
of the target system.  While the traditional state-space
reconstruction machinery is a good way to accomplish the task of
modeling the dynamics, it is problematic in real-time forecasting
because it generally requires input from and interpretation by a human
expert in order to work properly.  The strategy suggested in this
paper sidesteps that roadblock by using a reduced-order variant of
delay-coordinate embedding to build forecast models for nonlinear
dynamical systems.

Modern approaches to modeling a time series for the purposes of
forecasting arguably began with Yule's work on predicting the annual
number of sunspots\cite{Yule27} through what is now known as
autoregression.  Before this, time-series forecasting was done mostly
through simple global extrapolation\cite{weigend-book}.  Global linear
methods, of course, are rarely adequate when one is working with
nonlinear dynamical systems; rather, one needs to model the details of
the state-space dynamics in order to make accurate predictions.  The
usual first step in this process is to reconstruct that dynamical
structure from the observed data.  The state-space reconstruction
techniques proposed by Packard {\it et al.} \cite{packard80} in 1980
were a critical breakthrough in this regard.  In the most common
variant of this now-classic approach, one constructs a set of vectors
$\vec{x}_j\in\mathbb{R}^m$ where each coordinate is simply a
time-delayed element of the scalar time-series data $x_j$, {\it i.e.},
$\vec{x}_j=[x_j~x_{j-\tau} ~ \dots ~ x_{j-(m-1)\tau}]$ for $\tau>0$.
In 1981, Takens showed that this \emph{delay-coordinate embedding}
method provides a topologically correct representation of a nonlinear
dynamical system if a specific set of theoretical assumptions are
satisfied \cite{takens}; in 1991, Sauer {\it et al.}  extended this
discussion and relaxed some of the theoretical
restrictions~\cite{sauer91}.  This remains a highly active field of
research; see, for example, Abarbanel \cite{abarbanel} or Kantz \&
Schreiber \cite{kantz97} for surveys.
  
A large number of creative strategies have been developed for using
the state-space structure of a dynamical system to generate
predictions ({\it
e.g.},~\cite{casdagli-eubank92,weigend-book,Smith199250,sauer-delay,sugihara90,pikovsky86-sov}).
Perhaps the most simple of these is the ``Lorenz Method of Analogues''
(LMA), which is essentially nearest-neighbor
prediction~\cite{lorenz-analogues}.  Lorenz's original formulation of
this idea used the full system state space; this method was extended
to embedded dynamics by Pikovsky~\cite{pikovsky86-sov}, but is also
related to the prediction work of Sugihara \& May~\cite{sugihara90},
among others.  Even this simple strategy---which, as described in more
detail in Section~\ref{sec:forecastmodels}, builds predictions by
looking for the nearest neighbor of a given point and taking that
neighbor's observed path as the forecast---works quite well for
forecasting nonlinear dynamical systems.  LMA and similar methods have
been used successfully to forecast measles and chickenpox
outbreaks \cite{sugihara90}, marine phytoplankton
populations \cite{sugihara90}, performance dynamics of a running
computer~\cite{josh-IDA11,josh-IDA13,josh-pre}, the fluctuations in a
far-infrared laser\cite{sauer-delay,weigend-book}, and many more.

The reconstruction step that is necessary before these methods can be
applied to scalar time-series data, however, can be problematic.
Delay-coordinate embedding is a powerful piece of machinery, but
estimating values for its two free parameters, the time delay $\tau$
and the dimension $m$, is not trivial.  A large number of heuristics
have been proposed for this task ({\it e.g.}, \cite{Casdagli:1991a,
Gibson92,Buzugfilldeform,rosenstein94,liebert-wavering,
Liebert89,fraser-swinney,
kantz97,pecoraUnified,KBA92,Cao97Embed,Kugi96,Hegger:1999yq}),
but these methods, described in more detail in Section~\ref{sec:dce},
are computationally intensive and they require input from---and
interpretation by---a human expert.  This can be a real problem in a
prediction context: a millisecond-scale forecast is not useful if it
takes seconds or minutes to produce.  And it is even more of a problem
in nonstationary systems, since the reconstruction machinery is only
guaranteed to work for an infinitely long noise-free observation of
a \emph{single} dynamical system.  If effective forecast models are to
be constructed and applied in a manner that outpaces the dynamics of
the target system, then, one may not be able to use the full,
traditional delay-coordinate embedding machinery to reconstruct the
dynamics.

\begin{figure*}[ht!]
        \centering
        \begin{subfigure}[b]{\columnwidth}
                \includegraphics[width=\columnwidth]{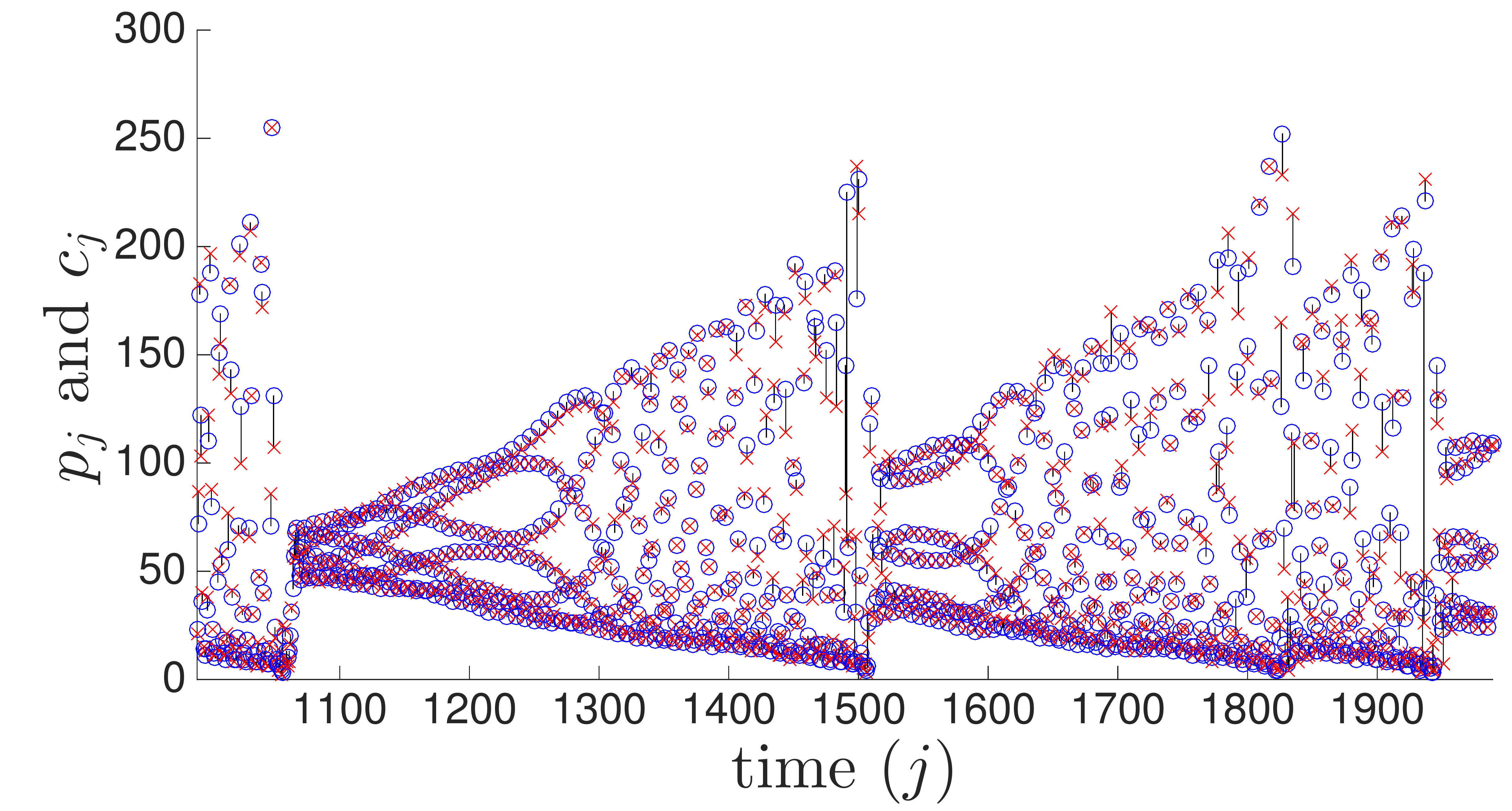}
                \caption{LMA in the full embedding space (``\fnnLMA'') }
                \label{fig:sfiA12D500tspred}
        \end{subfigure}%
        \begin{subfigure}[b]{\columnwidth}
                \includegraphics[width=\columnwidth]{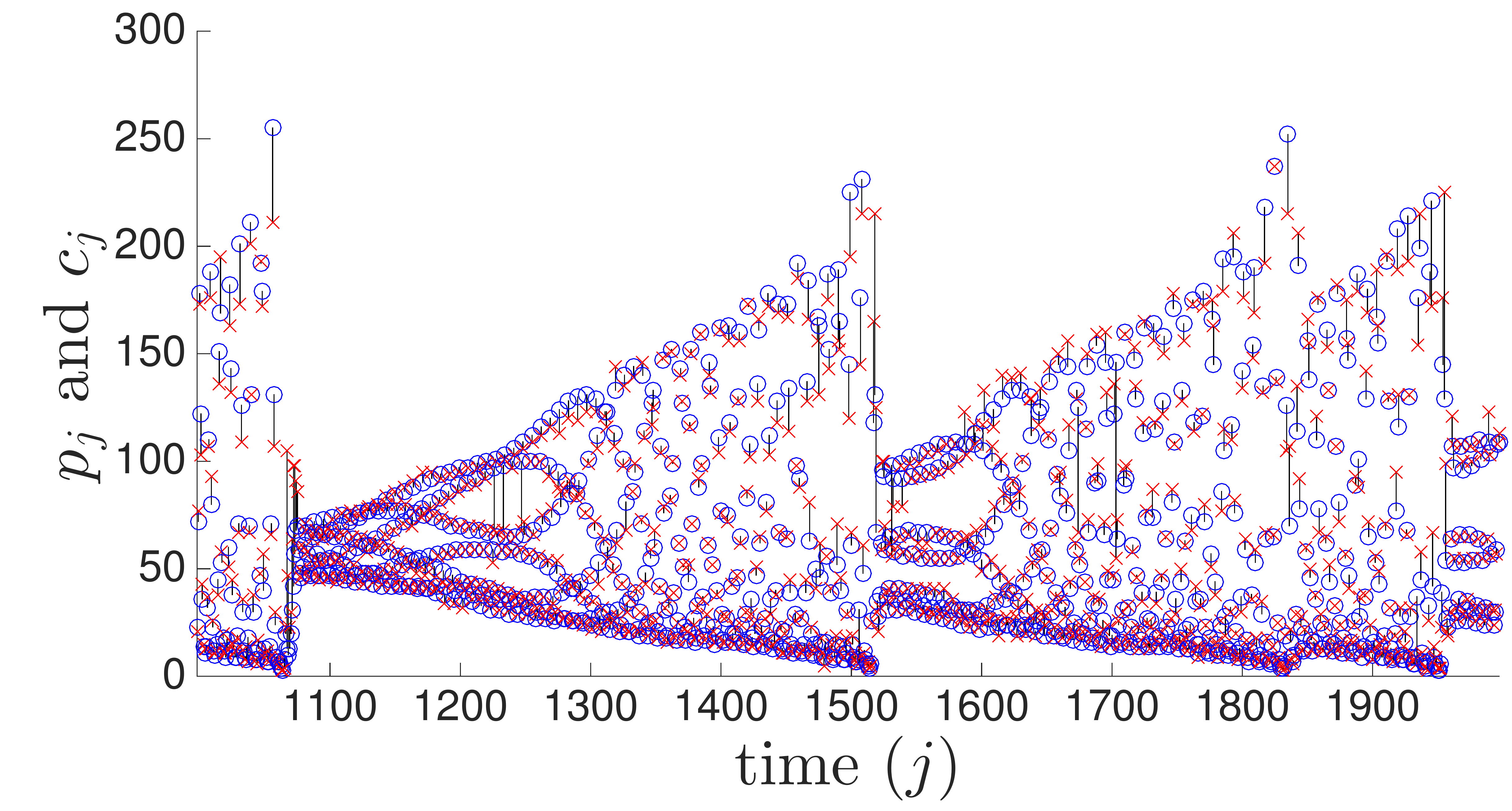}
                \caption{LMA in a two-dimensional embedding space (``\roLMA'')}
                \label{fig:sfiA2D500tspred}
        \end{subfigure}

\caption{Forecasts of SFI data set A using
Lorenz's method of analogues in (a) full and (b) 2D reconstructions of
the state-space dynamics.  Blue {\color{blue}o}s are the true
continuation $c_j$ of the time series and red {\color{red}x}s ($p_j$)
are the forecasts; black error bars are provided if there is a
discrepancy between the two.  Embedding parameter values were
estimated using standard techniques: the first minimum of the average
mutual information \cite{fraser-swinney} for the delay in both images
and the false near neighbor method of Kennel {\sl et al.}\cite{KBA92},
with a threshold of 20\%, for the dimension in the left-hand image.
Even though the $2D$ embedding used in (b) is not faithful to the
underlying topology, it enables successful forecasting of the time
series.}
%
%
\label{fig:projExample}
\end{figure*}

The goal of the work described in this paper was to sidestep that
problem by developing prediction strategies that work
in \emph{incomplete} embedding spaces.  The conjecture that forms the
basis for our work is that a full formal embedding, although mandatory
for detailed dynamical analysis, \emph{is not necessary for the
purposes of prediction}.  As a first step towards validating that
conjecture, we constructed two-embeddings from a number of different
time-series data sets, both simulated and experimental, and then built
forecast models in that space.  Sections~\ref{sec:projresults}
and~\ref{sec:time-scales} of this paper present and discuss those
results in some detail.  In short, we found that forecasts produced
using the Lorenz method of analogues on a two-dimensional
delay-coordinate embedding are roughly as accurate as---and often
even \emph{more} accurate than---forecasts produced by the same method
working in the full embedding space.  Figure~\ref{fig:projExample}
shows a quick example: a pair of forecasts of the so-called ``Dataset
A", a time series from a far-infrared laser from the Santa Fe
Institute prediction competition\cite{weigend-book}, generated with
LMA on full and $2D$ embeddings.

The main point of Figure~\ref{fig:projExample}---and the main claim of
this paper---concerns the similarity of the two panels.  The forecast
generated in the full embedding space is not much more accurate than
the one generated in the two-dimensional reconstruction.  The errors
between true and predicted values were 0.117 and 0.148, respectively,
as measured using the assessment procedure and error metric covered in
Section~\ref{sec:forecastmodels}; a better choice of $\tau$, as
described in Section~\ref{sec:time-scales}, brings the latter value
down to 0.119.  That is, even though the low-dimensional
reconstruction is not completely faithful to the underlying dynamics,
it appears to be good enough to support accurate forecast models of
nonlinear dynamics.  Both of these LMA-based forecasts, incidentally,
significantly outperformed traditional strategies like random-walk (by
a factor of $\approx 8.5$) and autoregressive integrated moving average (by a
factor of $\approx 6.5$).

The results in Sections~\ref{sec:projresults}
and~\ref{sec:time-scales} offer a deeper validation of the claim that
the full complexity (and effort) of the delay-coordinate `unfolding'
process may not be strictly necessary to the success of forecast
models of real-world nonlinear dynamical systems.  In effect, the
reduced-order modeling strategy that we propose is a kind of balance
of a tradeoff between the power of the state-space reconstruction
machinery and the effort required to use it.  Fixing $m=2$ effectively
avoids all of the in-depth, post-facto, data-dependent analysis that
is required to properly estimate a value for this parameter---which is
arguably the harder part of the process.  It also avoids the high
computational complexity that is involved in near-neighbor searches in
high-dimension state spaces---an essential step in almost any forecast
strategy.
%
Of course, there is still a free parameter: the delay $\tau$.  As
described in Section~\ref{sec:time-scales}, though, we believe that
good values for this parameter can be estimated `on the fly,' with
little to no human guidance, which would make this method both agile
and resilient.

No forecast model will be ideal for every task. In fact, as a
corollary of the undecidability of the halting
problem~\cite{halting-problem}, no single forecasting schema is ideal
for all noise-free deterministic signals~\cite{weigend-book}---let
alone all real-world time-series data sets.  We do not want to give
the impression that the strategy proposed here will be effective
for \emph{every} time series, but we do intend to show that it is
effective for a broad spectrum of signals.  Additionally, we want to
emphasize that it is a \emph{short}-term forecasting scheme.
Dimensional reduction is a double-edged sword; it enables on-the-fly
forecasting by eliminating a difficult estimation step, but it effects
a guaranteed information loss in the model.  This well-known
effect \cite{weigend-book} all but guarantees problems with accuracy
as prediction horizons are increased.  We explore this limitation in
Section~\ref{sec:roLMALorenz96}.
 

\section{Background and Methods}\label{sec:background}

\subsection{Delay-Coordinate Embedding}\label{sec:dce}

The process of collecting a time series $\{x_{j}\}_{j=1}^{N}$ from a
dynamical system (aka a ``trace'') is formally the evaluation of
an \emph{observation function} $h: \mathbb{X} \rightarrow \mathbb{R}$
at the true system state $\vec{y}(t_j)$ at time $t_j$ for $j=1,..,N$,
{\it i.e.}, $x_j = h(\vec{y}(t_j))$ for $j=1,\dots,N$ \cite{sauer91}.
Provided that the underlying dynamics $\Phi$ and the observation
function are both smooth and generic, Takens \cite{takens} proves that
the delay coordinate map:
\begin{equation}\label{eqn:takens}
F(h,\Phi,\tau,m)(\vec{y}(t_j)) = (  [x_j~x_{j-\tau} ~ \dots ~ x_{j-(m-1)\tau}]^T)
\end{equation}
from a $d$-dimensional smooth compact manifold $M$ to
$\mathbb{R}^{2d+1}$ is a diffeomorphism on $M$.  To assure topological
conjugacy, the proof requires that the embedding dimension $m$ must
be at least twice the dimension $d$ of the ambient space; a tighter
bound of $m>2d_{cap}$, the capacity dimension of the original dynamics,
was later established by Sauer {\sl et al.}\cite{sauer91}.
Operationalizing either of these theoretical constraints can be a real
challenge.  $d$ is not known and accurate $d_{cap}$ calculations are
not easy with experimental data.  The theoretical constraints on the
time delay are less stringent: $\tau$ must be greater than zero and
not a multiple of any orbit's period\cite{sauer91,takens}.  In
practice, however, the noisy and finite-precision nature of digital
data and floating-point arithmetic combine to make the choice of
$\tau$ much more delicate\cite{kantz97}.

As mentioned in the previous section, the forecast strategy proposed
in this paper sidesteps the challenge of estimating the embedding
dimension by fixing $m=2$.  Selection of a value for the remaining
free parameter, $\tau$, is still an issue, though.  There are dozens
of methods for this---{\it
e.g.}, \cite{Casdagli:1991a,Gibson92,Buzugfilldeform,rosenstein94,liebert-wavering,
Liebert89,fraser-swinney, kantz97}.  In this paper, we use the method
of \emph{mutual information}\cite{fraser-swinney, Liebert89}, in which
$\tau$ is chosen at the first minimum of the time-delayed mutual
information, calculated using the {\tt TISEAN}
package\cite{Hegger:1999yq}.  Fraser \& Swinney argue that this
minimizes the redundancy of the embedding coordinates, thereby
maximizing the information content of the overall delay
vector \cite{fraser-swinney}.  This choice is discussed and
empirically verified by Liebert and Schuster\cite{Liebert89}, although
agreement on this topic is by no means
universal\cite{hasson2008influence,martinerie92}.  And it is well
known that choice of $\tau$ is application- and system-specific: a
$\tau$ that works well for a Lyapunov exponent calculation may not
work well for other purposes\cite{Buzug92Comp,kantz97,rosenstein94}.
Indeed, as we show in Section~\ref{sec:varyingproj}, the $\tau$ value
suggested by mutual information calculations is rarely the optimal
choice for our reduced-order forecast strategy.  Even so, it is a
reasonable starting point, as it is the standard technique used in the
dynamical systems community.

The embedding dimension $m$ is not a parameter in our reduced-order
forecasting strategy.  Since full embeddings are the point of
departure for the central premise of this paper, however---and the
point of comparison for our results---we briefly review methods used
to estimate values for that parameter.  As in the case of $\tau$, a
number of creative strategies for doing so have been developed over
the past few
decades \cite{KBA92,Cao97Embed,liebert-wavering,Kugi96,Buzugfilldeform,KBA92,Hegger:1999yq,kantz97}.
Most of these are based in some way on minimizing the number of false
crossings that are caused by projection.  In this paper, we use the
false near neighbor (FNN) approach of Kennel {\sl et al.}\cite{KBA92},
calculated using {\tt TISEAN}\cite{Hegger:1999yq} with a $\approx20\%$
threshold on the percentage of neighbors.  (Whenever we refer to a
``full'' embedding, as in the discussion of
Figure~\ref{fig:projExample}, we mean that the $m$ value was estimated
in this fashion and the $\tau$ value was chosen at the first minimum
of the mutual information, as described in the previous paragraph.)
Again, no heuristic method is perfect, but FNN is arguably the most
widely used method for estimating $m$, and thus is useful for the
purposes of comparison.

Finally, it should be noted that there is an alternative view that one
should estimate $\tau$ and $m$ together, not
separately\cite{Kugi96,McNames98anearest,rosenstein94,pecoraUnified}.

\subsection{Lorenz Method of Analogues}\label{sec:forecastmodels}

As mentioned in Section~\ref{sec:intro}, the dynamical systems
community has developed a number of methods that capture and use
state-space structure to create forecasts.  Since the goal of this
paper is to offer a proof of concept of the notion that incomplete
embeddings could give these kinds of methods enough traction to
generate useful predictions, we chose one of the oldest and simplest
members of that family: Lorenz's method of analogues (LMA).  In future
work, we will explore reduced-order versions of other forecast
strategies.

To apply LMA to a scalar time-series data
set $\{x_j\}_{j=1}^n$, one begins by performing a delay-coordinate
embedding using one or more of the heuristics presented in
Section~\ref{sec:dce} to choose $m$ and $\tau$.  This  produces a
trajectory of the form:
 $$\{\vec{x}_j=[x_j~x_{j-\tau} ~ \dots ~ x_{j-(m-1)\tau}]^T \}_{j=1-(m-1)\tau}^{n}$$ 
Forecasting the next point in the time series, $x_{n+1}$, amounts to
reconstructing the next delay vector $\vec{x}_{n+1}$ in the
trajectory.  Note that, by the form of delay-coordinate vectors, all
but the first coordinate of $\vec{x}_{n+1}$ are known.  To choose the
first coordinate, one finds the nearest neighbor of $\vec{x}_{n}$ in
the reconstructed space---namely $\vec{x}_{j(1,m)}$---and maps that
vector forward using the delay map, obtaining 
$$ \vec{x}_{j(1,m)+1}=[x_{j(1,m)+1}~x_{j(1,m)+1-\tau} ~ 
\dots ~ x_{j(1,m)+1-(m-1)\tau}]^T$$
Using the neighbor's image, one defines $$\vec{p}_{n+1} =
[x_{j(1,m)+1}~x_{n+1-\tau} ~ \dots ~ x_{n+1-(m-1)\tau}]^T$$ LMA
defines the forecast of $x_{n+1}$ as $p_{n+1}=x_{j(1,m)+1}$.  If
performing multi-step forecasts, one appends the new
delay vector $$\vec{p}_{n+1}=[x_{j(1,m)+1}~x_{n+1-\tau} ~ \dots ~
x_{n+1-(m-1)\tau}]^T$$ to the end of the trajectory and repeats this
process as needed.

Many more-complicated variants of this algorithm have appeared in the
literature ({\it
e.g.},~\cite{weigend-book,casdagli-eubank92,Smith199250,sugihara90}),
most of which involve building some flavor of local-linear models
around each delay vector and then using it to make the prediction of
the next point.  Here, we use the basic version, in two ways:
first---as a baseline for comparison purposes---on a ``full''
embedding of each time series, with $m$ chosen using the false near
neighbor method of Kennel {\sl et al.}\cite{KBA92} on that data;
second, with $m=2$.  In the rest of this paper, we will refer to these
as \fnnLMA and \roLMA, respectively.  The experiments reported in
Section~\ref{sec:projresults} use the same $\tau$ value for
both \fnnLMA and \roLMA, choosing it at the first minimum of the
time-delayed mutual information of the time
series\cite{fraser-swinney}.  In Section~\ref{sec:time-scales}, we
explore the effects of varying $\tau$ on the accuracy of both methods.

\subsection{Assessing Forecast Accuracy}
\label{sec:accuracy}

To evaluate \roLMA and compare it to \fnnLMA, we calculate a figure of
merit in the following way.  We split each $N$-point time series into
two pieces: the first 90\%, referred to as the ``initial training"
signal and denoted $\{x_j\}_{j=1}^{n}$, and the last 10\%, known as
the ``test" signal $\{c_j\}_{j=n+1}^{(k+n+1)=N}$.  We use the initial
training signal to build the model, following the procedures described
in the previous section.  We use that model to generate a prediction
$p_{n+1}$ of the value of $x_{n+1}$, then compare $p_{n+1}$ to the
true continuation, $c_{n+1}$.  The initial investigations that are
reported in Section~\ref{sec:projresults} involve ``one-step'' models,
which are rebuilt after each step, out to the end of the test signal,
using $\{c_{n+1}\}\cup\{x_j\}_{j=1}^{n}$.  In
Section~\ref{sec:time-scales}, we extend this conversation to longer
prediction horizons.

As a numerical measure of prediction accuracy, we calculate the mean
absolute scaled error ($MASE$)\cite{MASE} between the true and predicted
signals:
$$MASE = \sum_{j=n+1}^{k+n+1}\frac{|p_j-c_j|
}{\frac{k}{n-1}\sum^n_{i=2}|x_{i}-x_{i-1}|}$$
$MASE$ is a normalized measure: the scaling term in the denominator is
the average in-sample forecast error for a random-walk
prediction---which uses the previous value in the observed signal as
the forecast---calculated over the initial training signal
$\{x_j\}^n_{j=1}$.  That is, $MASE<1$ means that the prediction error
in question was, on the average, smaller than the in-sample error of a
random-walk forecast on the same data.  Analogously, $MASE>1$ means
that the corresponding prediction method did \emph{worse}, on average,
than the random-walk method.  The $MASE$ value of 0.117 for
Figure~\ref{fig:projExample}, for instance, means that the \fnnLMA
forecast of the SFI data set A was $\frac{1}{0.117}$ times better than
a random-walk forecast of the initial training portion of same signal.

While its comparative nature may seem odd, this error metric allows
for fair comparison across varying methods, prediction horizons, and
signal scales, making it a standard error measure in the forecasting
literature---and a good choice for the study described in the
following sections, which involve a number of very different signals.

\section{Prediction In Projection}
\label{sec:projresults}

In this section, we demonstrate that the accuracies of forecasts
produced by \roLMA---Lorenz's method of analogues, operating on a
two-dimensional embedding of a trajectory from a dynamical
system---are similar to and often better than forecasts produced
by \fnnLMA, which operates on a full reconstruction of the same
dynamics.  While the brief example in Section~\ref{sec:intro} is a
useful first validation of that statement, it does not support the
kind of exploration that is necessary to properly evaluate a new
forecast method, especially one that violates the basic tenets of
delay-coordinate embedding.  The SFI data set A is a single trace from
a single system.  We want to show that \roLMA is comparable to or
better than \fnnLMA for a \emph{range} of systems and parameter
values---and to repeat each experiment for a number of different
trajectories from each system.  To this end, we studied two dynamical
systems, one simulated and one real: the Lorenz-96 model and sensor
data from a laboratory experiment on computer performance dynamics.

\subsection{A Synthetic Example: Lorenz-96}\label{sec:roLMALorenz96}
 
The Lorenz-96 model\cite{lorenz96Model} is a set of $K$ first-order
differential equations relating the $K$ state variables
$\xi_1\dots\xi_K$:
\begin{equation}\label{eq:lorenz96}
\dot{\xi}_k= (\xi_{k+1}-\xi_{k-2})(\xi_{k-1})-\xi_k + F
\end{equation}
for $k=1,\dots,K$, where $F\in \mathbb{R}$ is a constant forcing term
that is independent of $k$.  In this model, each $\xi_k$ is some
atmospheric quantity (such as temperature or vorticity) at a discrete
location on a periodic lattice representing a latitude circle of the
earth \cite{KarimiL96}.  This model exhibits a wide range of dynamical
behavior---everything from fixed points and periodic attractors to low-
and high-dimensional chaos\cite{KarimiL96}---making it an ideal test
case for our purposes.

We performed two sets of forecasting experiments with traces from the
Lorenz-96 model: one with $K=22$ and the other with $K=47$.  Both
experiments used constant forcing values of $F=5$.  These choices
yield chaotic trajectories with low and high Kaplan-Yorke (Lyapunov)
dimension\cite{kydimension}: $d_{KY}\lessapprox3$ for the $K=22$
dynamics and $d_{KY}\approx19$ for $K=47$~\cite{KarimiL96}. Following
standard practice~\cite{KarimiL96}, we enforced periodic boundary
conditions and solved equation~(\ref{eq:lorenz96}) from several
randomly chosen initial conditions using a standard fourth-order
Runge-Kutta integrator for 60,000 steps with a step size of
$\tfrac{1}{64}$ normalized time units.  We then discarded the first
10,000 points of that trajectory in order to eliminate transient
behavior.  Finally, we created scalar time-series traces by
individually ``observing'' each of the $K$ state variables of the
trajectory: {\it i.e.,} $h_i(\xi_i(t_j)) = x_{j,i}$ for
$j\in\{10,000,\dots,60,000\}$ and for $i\in \{1,\dots,K\}$.  We
repeated all of this from a number of different initial
conditions---seven for the $K=47$ Lorenz-96 system and 15 for the
$K=22$ case---producing a total of 659 traces for our forecasting
study.  For each of these, we used the procedures outlined in
Section~\ref{sec:dce} to estimate values for the free parameters of
the embedding process, obtaining $m=8$ and $\tau=26$ for all traces in
the $K=22$ case, and $m=10$ and $\tau=31$ for the $K=47$
traces\footnote{It has been shown\cite{sprottBook} that
$d_{KY} \approx d_{cap}$ for typical chaotic systems.  This suggests
that embeddings of the $K=22$ and $K=47$ time series would require
$m\gtrapprox6$ and $m\gtrapprox 38$, respectively.  The values
suggested by the false-near neighbor method for the $K=22$ traces are
in line with this, but the $K=47$ FNN values are far smaller than $2
d_{KY}$.}.

\begin{figure}[ht!]
        \centering
        \begin{subfigure}[b]{\columnwidth}
                \includegraphics[width=\columnwidth]{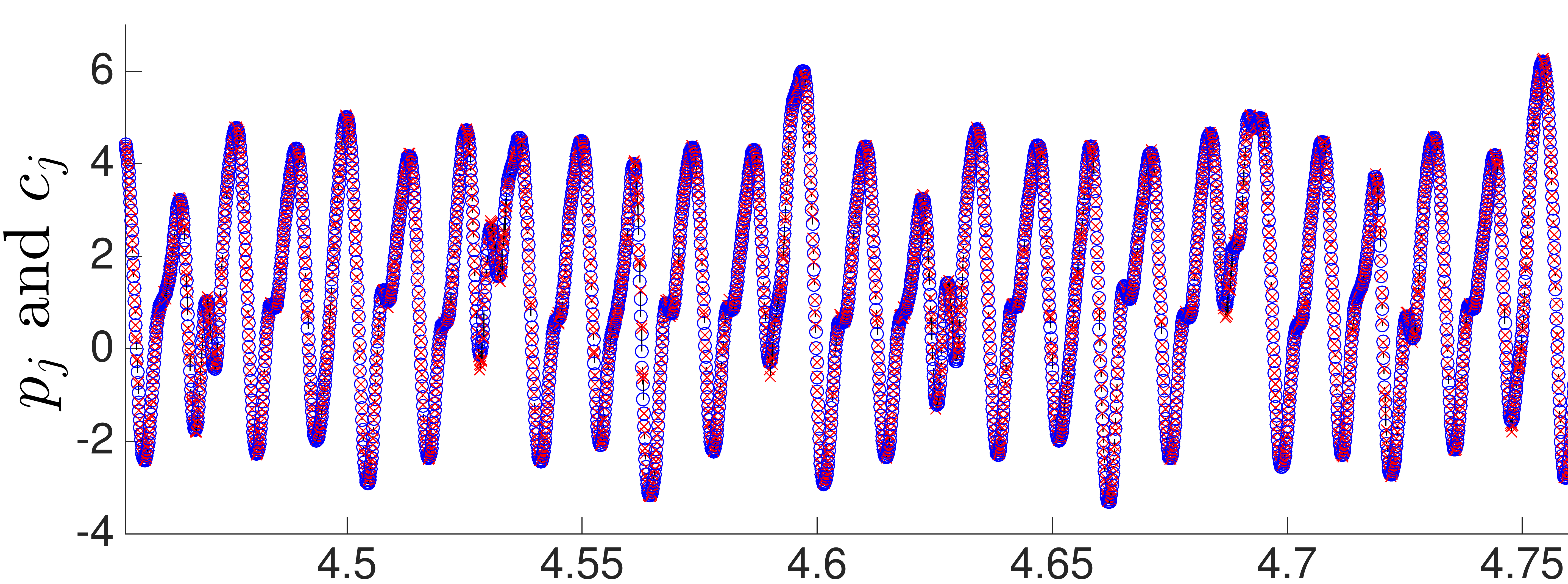}

                \caption{5,000-point forecast using the reduced-order
                forecast method \roLMA. Blue circles and red
                {\color{red}$\times$}s are the true and predicted
                values, respectively; vertical bars show where these
                values differ.}

                \label{fig:K22m2k1per10tspred}
        \end{subfigure}%
        ~ 
          
        \begin{subfigure}[b]{0.49\columnwidth}
                \includegraphics[width=\columnwidth]{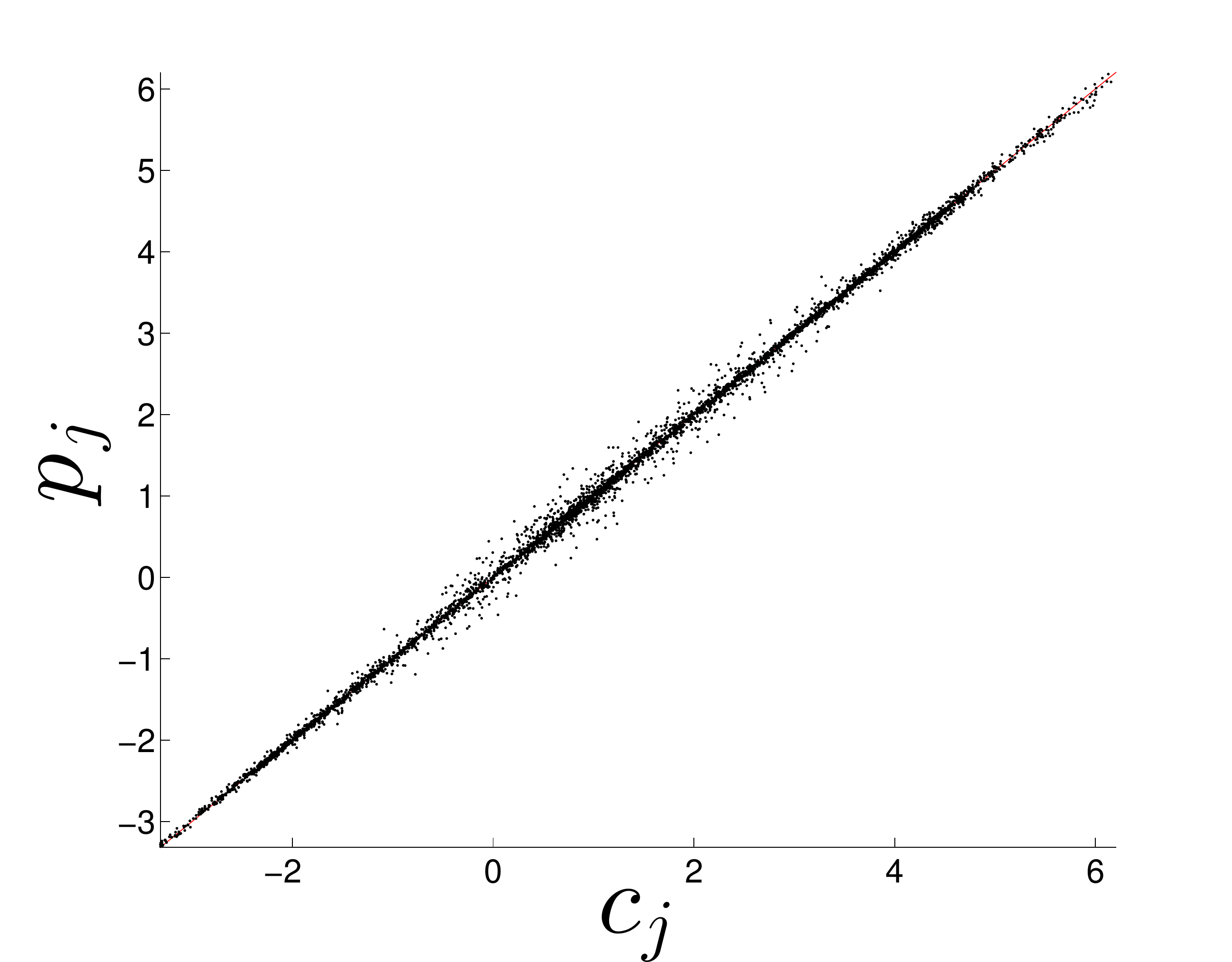}
                \caption{\roLMA forecast}
                \label{fig:K22m2k1per10pjcjpred}
        \end{subfigure}
                \begin{subfigure}[b]{0.49\columnwidth}
                \includegraphics[width=\columnwidth]{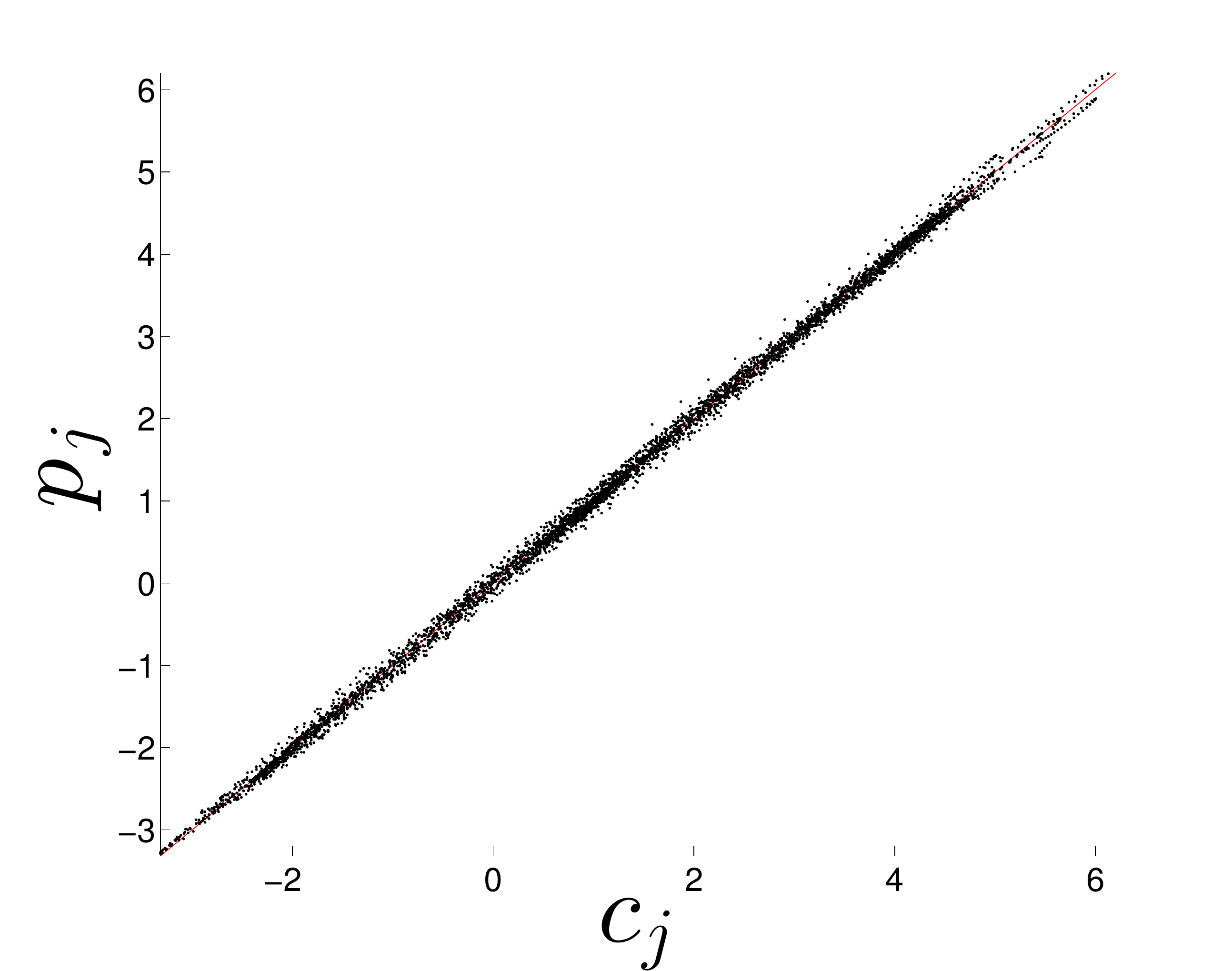}
                \caption{\fnnLMA forecast}
                \label{fig:K22m8k1per10pjcjpred}
        \end{subfigure}

\caption{\roLMA and \fnnLMA forecasts of a representative trace from the
Lorenz-96 system with $K=22$ and $F=5$.  Top: a time-domain plot of
the first 5,000 points of the \roLMA forecast.  Bottom: the predicted
($p_j$) vs true ($c_j$) values for forecasts of that trace generated
by (b) \roLMA and (c) \fnnLMA.  On such a plot, a perfect prediction
would lie on the diagonal.  The $MASE$ scores of the forecasts in (b)
and (c) were 0.392 and 0.461, respectively.
}\label{fig:K22predictions}
\end{figure}

For the $K=22$ dynamics, both \roLMA and \fnnLMA worked quite well.
See Figure~\ref{fig:K22predictions}(a) for a time-domain plot of
an \roLMA forecast of a representative trace from this system and
Figures~\ref{fig:K22predictions}(b) and (c) for graphical
representations of the forecast accuracy on that trace for both
methods.
The diagonal structure on the $p_j$ vs. $c_j$ plots in the Figure
indicates that both of these LMA-based methods performed very well on
this trace.  More importantly---from the standpoint of evaluation of
our primary claim---the $MASE$ scores of \roLMA and \fnnLMA forecasts,
computed following the procedures described in
Section~\ref{sec:accuracy}, were $0.391\pm0.016$ and $0.441 \pm
0.033$, respectively, across the 330 traces at this parameter value.
That is, the LMA forecasting strategy worked \emph{better} on a
two-dimensional embedding of these dynamics than on a full embedding,
and by a statistically significant margin.  This is somewhat
startling, given that the two-embedding is not topologically correct.
Clearly, though, it captures \emph{enough} structure to allow LMA to
generate good predictions.  And \roLMA's reduced-order nature may
actually mitigate the impact of noise effects, simply because a single
noisy point in a scalar time series affects $m$ of the points in an
$m$-embedding.
\label{page:mitigate}
Of course, $\tau$ choice, information content, and/or data length
could also be at work in these results; these concerns are addressed
in Sections~\ref{sec:time-scales} and~\ref{sec:concl}.

The $K=47$ case is a slightly different story: \roLMA still
outperformed \fnnLMA, but not by a statistically significant margin.
The $MASE$ scores across all 329 traces were $0.985 \pm 0.047$ and
$1.007 \pm 0.043$ for \roLMA and \fnnLMA, respectively.  In view of
the higher complexity of the state-space structure of the $K=47$
version of the Lorenz-96 system, the overall increase in $MASE$ scores over
the $K=22$ case makes sense.
Recall that $d_{KY}$ is far higher for the $K=47$ case: this attractor
fills more of the state space and has many more dimensions that are
associated with positive Lyapunov exponents.  This has obvious
implications for predictability.  It could very well be the case that
more data are necessary to reconstruct these dynamics, but our
experiments held the length of the training set constant across all of
the Lorenz-96 experiments.  These issues are described at more length
in Section~\ref{sec:concl}.  Regardless, it is encouraging that the
reduced-order forecasting method still performs as well as the version
that works with a fully `unfolded' embedding.

\subsection{Experimental Data: Computer Performance Dynamics}
\label{sec:compPerfProj}

Validation with synthetic data are an important first step in
evaluating any new forecast strategy, but experimental time-series
data are the acid test if one is interested in real-world
applications.  Our second set of tests of \roLMA, and comparisons of
its accuracy to that of \fnnLMA, involved data from a laboratory
experiment on computer performance dynamics.  Like Lorenz-96, this
system has been shown to exhibit a range of interesting deterministic
dynamical behavior, from periodic orbits to low- and high-dimensional
chaos\cite{zach-IDA10,mytkowicz09}, making it a good test case for
this paper.  It also has important practical implications; these
dynamics, which arise from the deterministic, nonlinear interactions
between the hardware and the software, have profound effects on
execution time and memory use.
%
%

Collecting observations of the performance of a running computer is
not trivial.  We used the {\tt libpfm4} library, via PAPI (Performance
Application Programming Interface) 5.2~\cite{papi}, to stop program
execution at 100,000-instruction intervals---the unit of time in these
experiments---and read the contents of the microprocessor's onboard
hardware performance monitors, which we had programmed to observe
important attributes of the system's dynamics.  See \cite{todd-phd}
for an in-depth description of this experimental setup.  The signals
that are produced by this apparatus are scalar time-series
measurements of system metrics like processor efficiency ({\it e.g.,}
\ipc, which measures how many instructions are being executed, on 
the average, in each clock cycle) or memory usage ({\it e.g.,} how
often the processor had to access the main memory during the
measurement interval).

\begin{figure}[ht!]
        \centering
        \begin{subfigure}[b]{\columnwidth}
\includegraphics[width=\columnwidth]{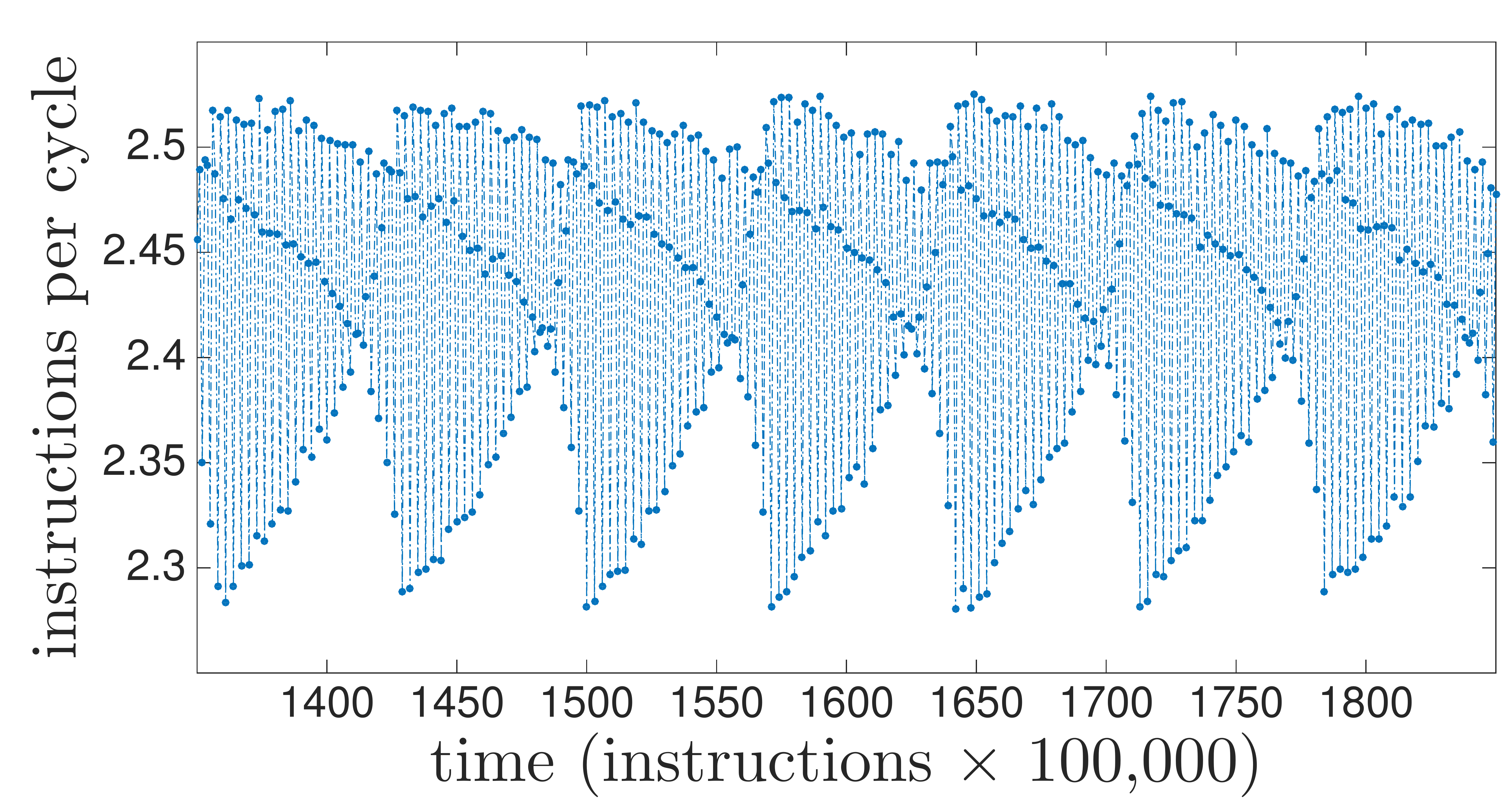}
                  \caption{A short segment of a trace of the
instructions executed per cycle (\ipc) during the execution of \col}

        \end{subfigure}
        \begin{subfigure}[b]{\columnwidth}
  \includegraphics[width=\columnwidth]{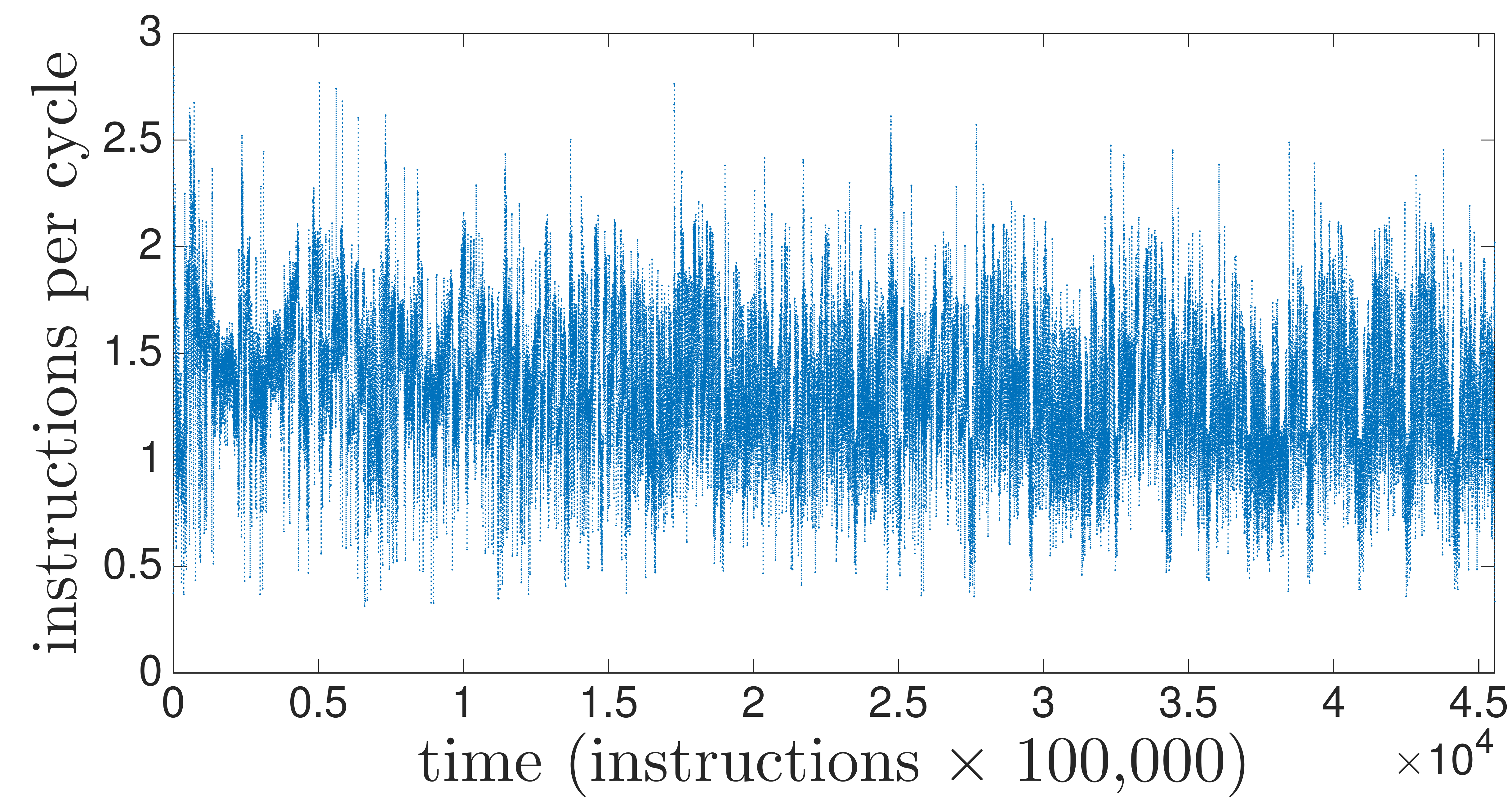}
                  \caption{A trace of the
instructions executed per cycle (\ipc) during the execution of \gcc}
        \end{subfigure}
\caption{Time-series data from a computer performance experiment:
processor load traces, in the form of instructions executed per cycle
(\ipc) of (a) a simple program (\col) that repeatedly initializes a
256 $\times$ 256 matrix and (b) a complex program (\gcc) from the SPEC
benchmark suite.  Each point is the average \ipc over a 100,000
instruction period. }
\label{fig:computerdata}
\end{figure}

We have tested \roLMA on traces of many different processor and memory
performance metrics gathered during the execution of a variety of
programs on several different computers.  Here, for conciseness, we
focus on \emph{processor} performance traces from two different
programs, one simple and one complex, running on the same Intel
i7-based computer.  Figure~\ref{fig:computerdata}(a) shows a small
segment of an \ipc time series gathered from that computer as it
executed a four-line C program (\col) that repeatedly initializes a
$256 \times 256$ matrix in column-major order.
On the scale of this figure, these dynamics appear to be largely
periodic, but they are actually chaotic \cite{mytkowicz09}.  The
bottom panel in Figure~\ref{fig:computerdata} shows a processor
efficiency trace from a much more complex program: the
\gcc compiler from the SPEC 2006CPU benchmark suite\cite{spec2006}.  

Since computer performance dynamics result from a composition of
hardware and software, these two programs represent two different
dynamical systems, even though they are running on the same computer.
The dynamical differences are visually apparent from the traces in
Figure~\ref{fig:computerdata}; they are mathematically apparent from
nonlinear time-series analysis of embeddings of those
data\cite{mytkowicz09}, as well as in calculations of the information
content of the two signals.  Among other things, \gcc has much less
predictive structure than \col and is thus much harder to
forecast\cite{josh-pre}.  These attributes make this a useful pair of
experiments for an exploration of the utility of reduced-order
forecasting.

\begin{figure}[htbp]
  \centering
      \begin{subfigure}{0.49\columnwidth}
    \includegraphics[width=\columnwidth]{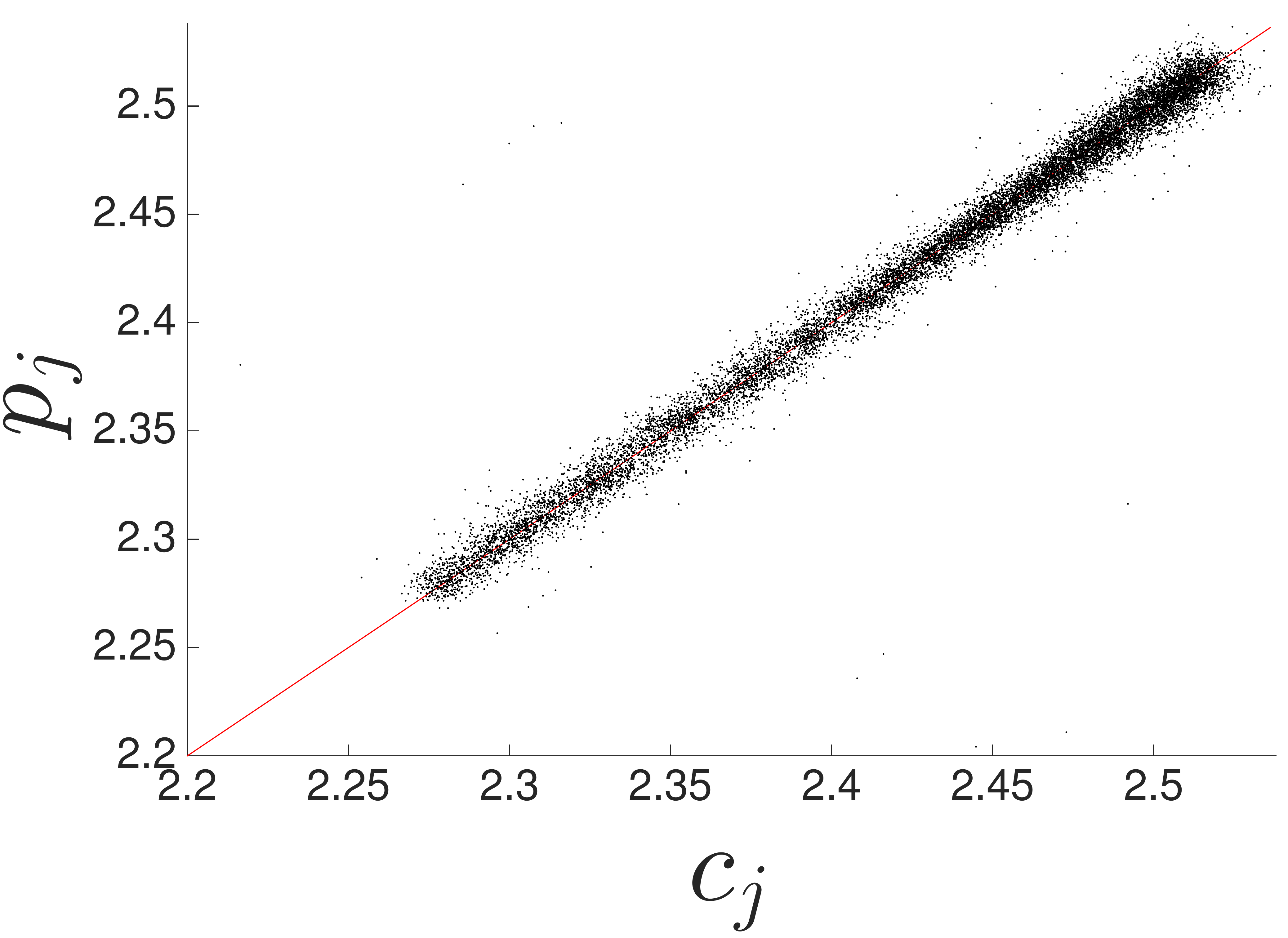}
    \caption{\fnnLMA on \col}
    \label{fig:colfnnLMA}
  \end{subfigure}
      \begin{subfigure}{0.49\columnwidth}
    \includegraphics[width=\columnwidth]{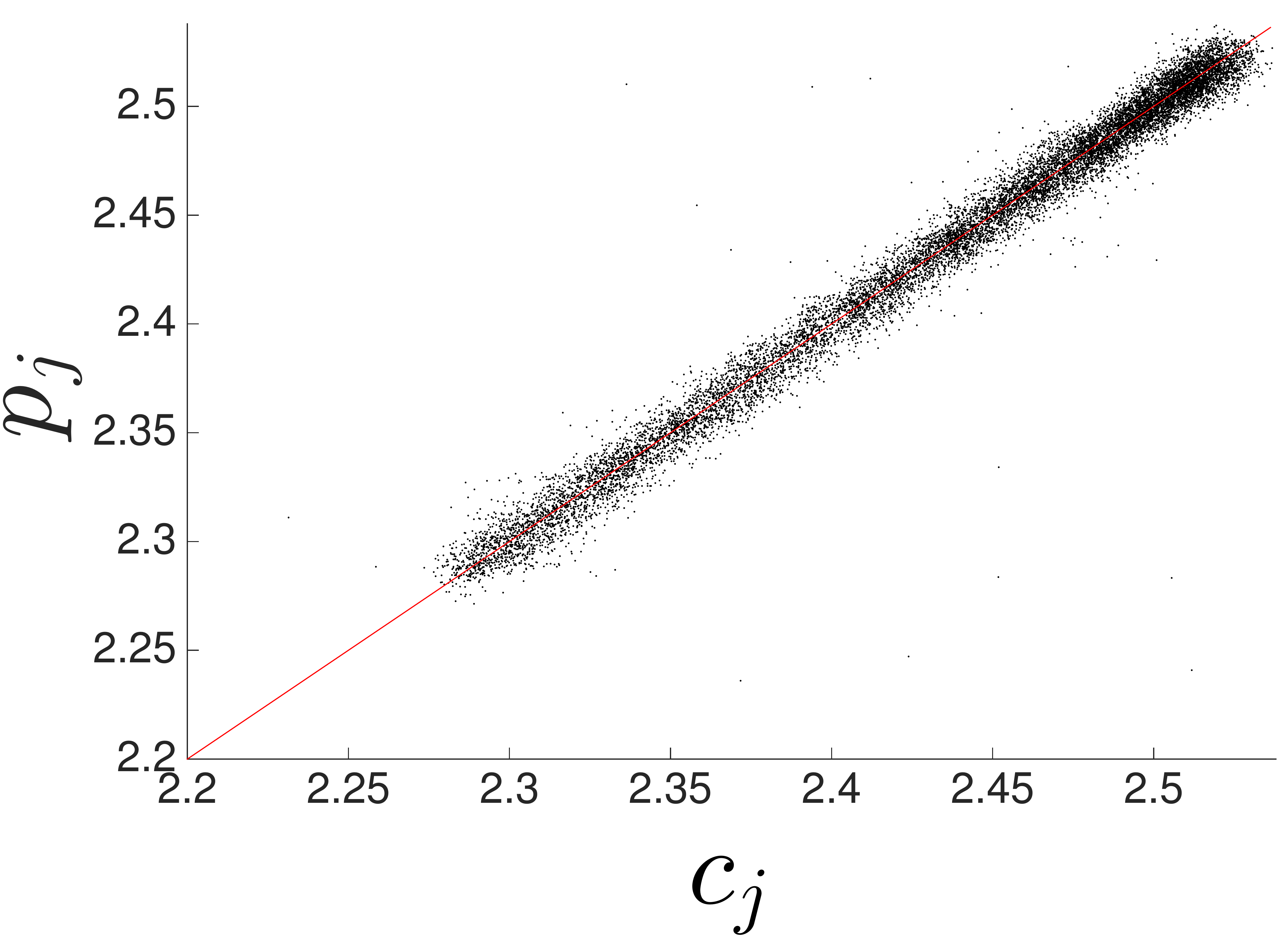}
    \caption{\roLMA on \col}
    \label{fig:colroLMA}
  \end{subfigure}
  \\
      \begin{subfigure}{0.49\columnwidth}
    \includegraphics[width=\columnwidth]{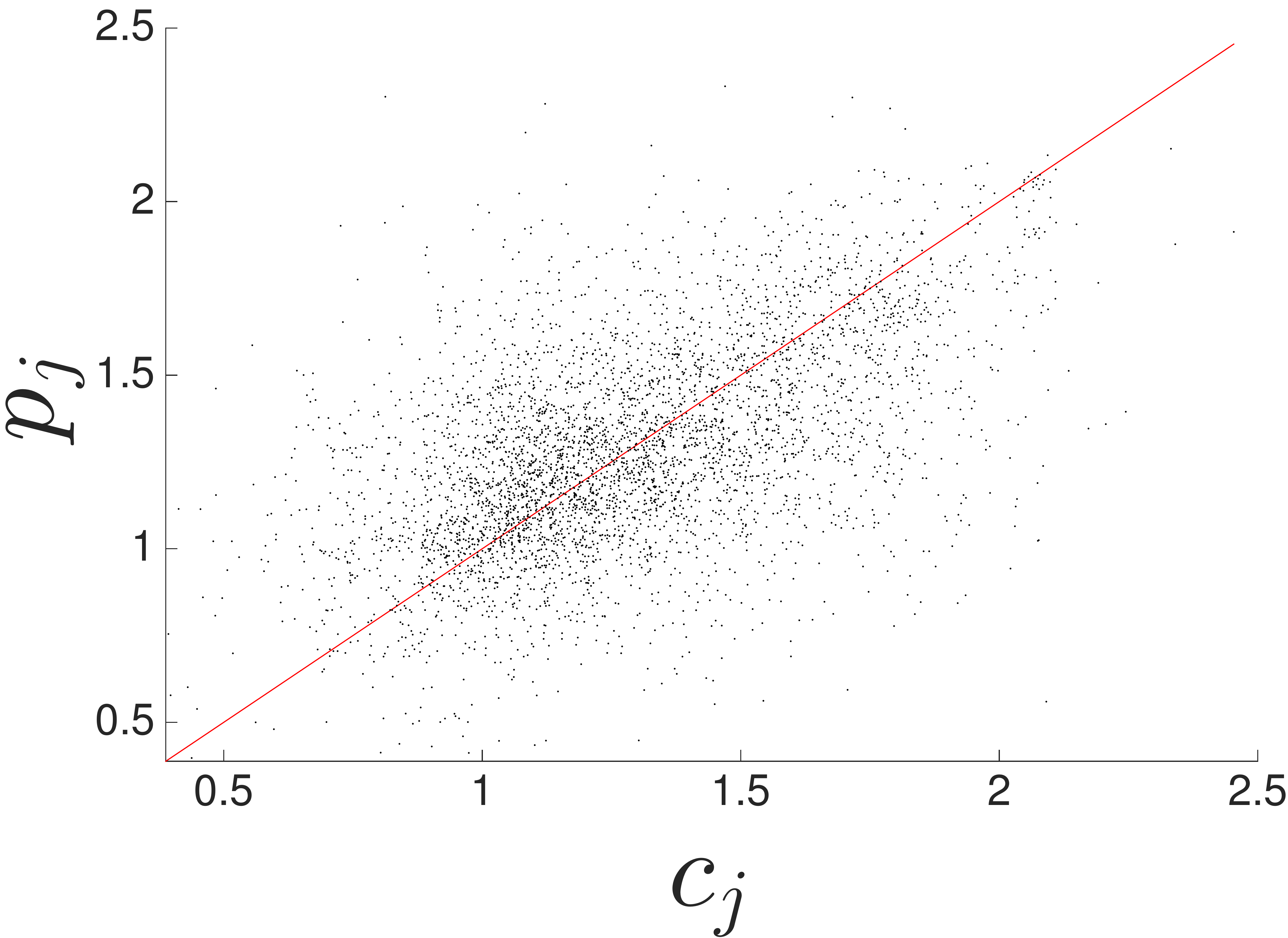}
    \caption{\fnnLMA on \gcc}
    \label{fig:gccfnnLMA}
  \end{subfigure}  
   \begin{subfigure}{0.49\columnwidth}
    \includegraphics[width=\columnwidth]{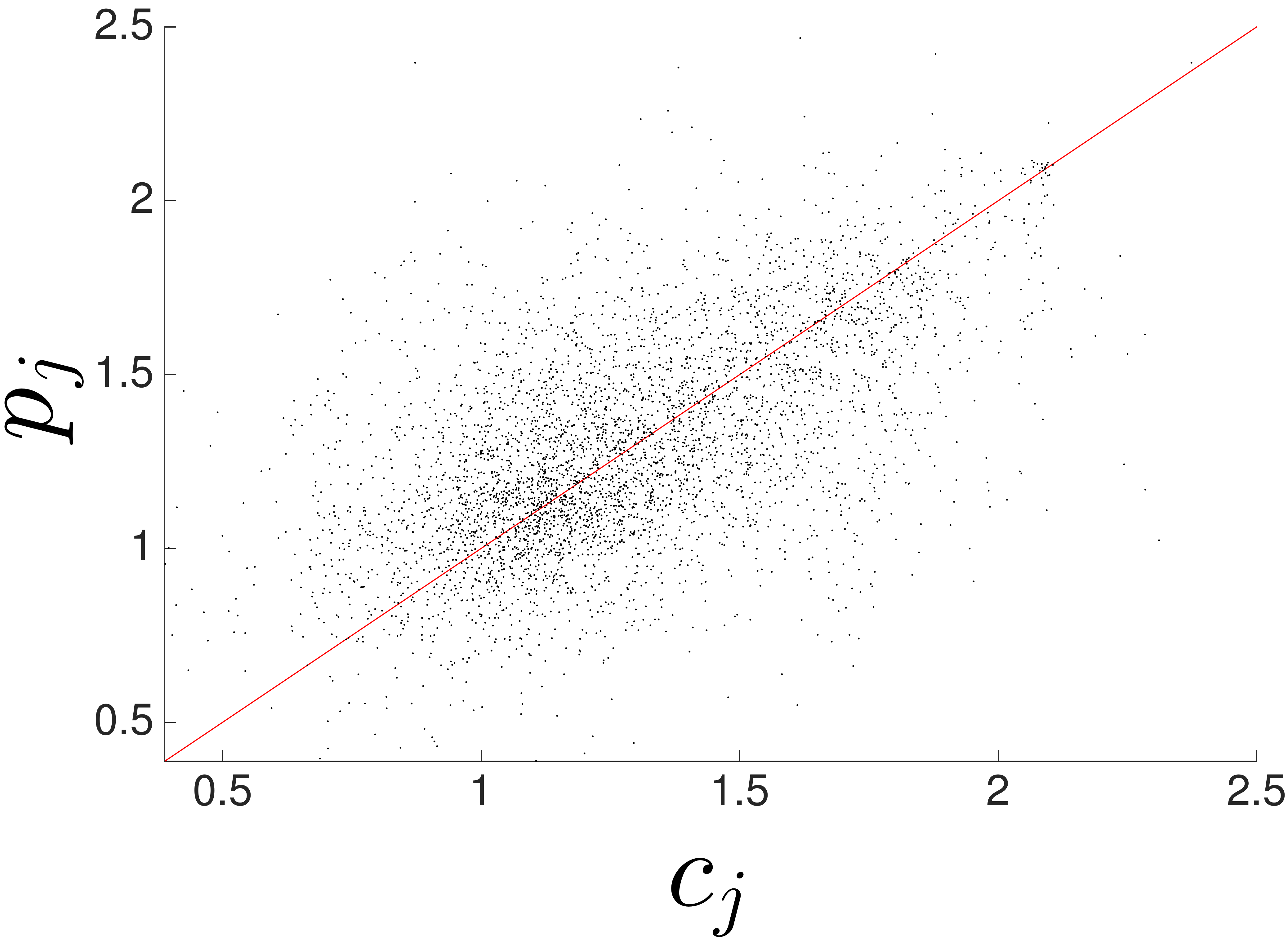}
    \caption{\roLMA on \gcc}
    \label{fig:gccroLMA}
  \end{subfigure}  
    \caption{Predicted ($p_j$) versus true values ($c_j$) for
     forecasts of \col and \gcc generated with \fnnLMA and \roLMA.}
\label{fig:forecast-example}
\end{figure}

For statistical validation, we collected 15 performance traces from
the computer as it ran each program, calculated embedding parameter
values as described in Section~\ref{sec:dce}, and generated forecasts
of each trace using \roLMA and \fnnLMA.
%
Figure~\ref{fig:forecast-example} shows $p_j$ vs. $c_j$ plots for
these forecasts.
The diagonal structure on the top two plots indicates that both \roLMA
and \fnnLMA performed well on the \col traces.  The $MASE$ scores
across all 15 trials in this set of experiments were $0.050 \pm 0.002$
and $0.063 \pm 0.003$, respectively---{\it i.e.,} \roLMA's accuracy
was somewhat worse than that of \fnnLMA.  For \gcc, however, \roLMA
appears to be somewhat more accurate: $1.488 \pm 0.016$
versus \fnnLMA's $1.530 \pm 0.021$.  Note that the \gcc $MASE$ scores
were higher for both forecast methods than on \col, simply because
the \gcc signal contains less predictive structure\cite{josh-pre}.
This actually makes the comparison somewhat problematic, as discussed
at more length on page~\pageref{page:suspect}.

Table~\ref{tab:error} summarizes results from all of the experiments
presented so far.  Overall, the results on the computer-performance
data are consistent with those that we obtained with the Lorenz-96
example in the previous section: prediction accuracies of \roLMA
and \fnnLMA were quite similar on all traces, despite the former's use
of an incomplete embedding.  This amounts to a validation of the
conjecture on which this paper is based.  And in both numerical and
experimental examples, \roLMA actually \emph{outperformed} \fnnLMA on
the more-complex traces (\gcc, $K=47$).  We believe that this is due
to the noise mitigation that is naturally effected by a
lower-dimensional embedding (cf., page~\pageref{page:mitigate}).
Finally, we found that it was possible to improve the performance of
both \roLMA and \fnnLMA on all four of these dynamical systems by
adjusting the free parameter, $\tau$.  It is to this issue that we
turn next; following that, we address the issue of prediction horizon.

\begin{table}[tb!]
\caption{$MASE$ scores for both forecast methods and all four sets of experiments.}
  \begin{center}
  \begin{tabular}{lccc}
  \hline\hline Signal & \fnnLMA & \roLMA & Trials\\ 
  \hline
Lorenz-96 $K=22$ ~~ & $0.441 \pm 0.033$ & $0.391\pm0.016$ & 330\\
Lorenz-96 $K=47$ & $1.007 \pm 0.043$ & $0.985 \pm 0.047$& 329\\
 \col           & $ 0.050 \pm0.002  $ &$0.063\pm0.003$ & 15\\

\gcc           & $ 1.5297\pm 0.0214$ &$1.488\pm0.016$ & 15 \\
%
%
%
%
  \hline\hline
  \end{tabular}
  \end{center}
 \label{tab:error}
  \end{table}%

\section{Time Scales, Parameters, and Prediction Horizons}
\label{sec:time-scales}


\subsubsection{The $\tau$ parameter}\label{sec:varyingproj}

The embedding theorems require only that $\tau$ be greater than zero
and not a multiple of any period of the dynamics.  In practice,
however, $\tau$ can play a critical role in the success of
delay-coordinate embedding---and any nonlinear time-series analysis
that follows\cite{fraser-swinney,kantz97,rosenstein94}. It should not
be surprising, then, that $\tau$ might affect the accuracy of an
LMA-based method that uses the structure of an embedding to make
forecasts.

\begin{figure}
        \centering

        ~ 
          
        \begin{subfigure}[b]{\columnwidth}
                \includegraphics[width=\columnwidth]{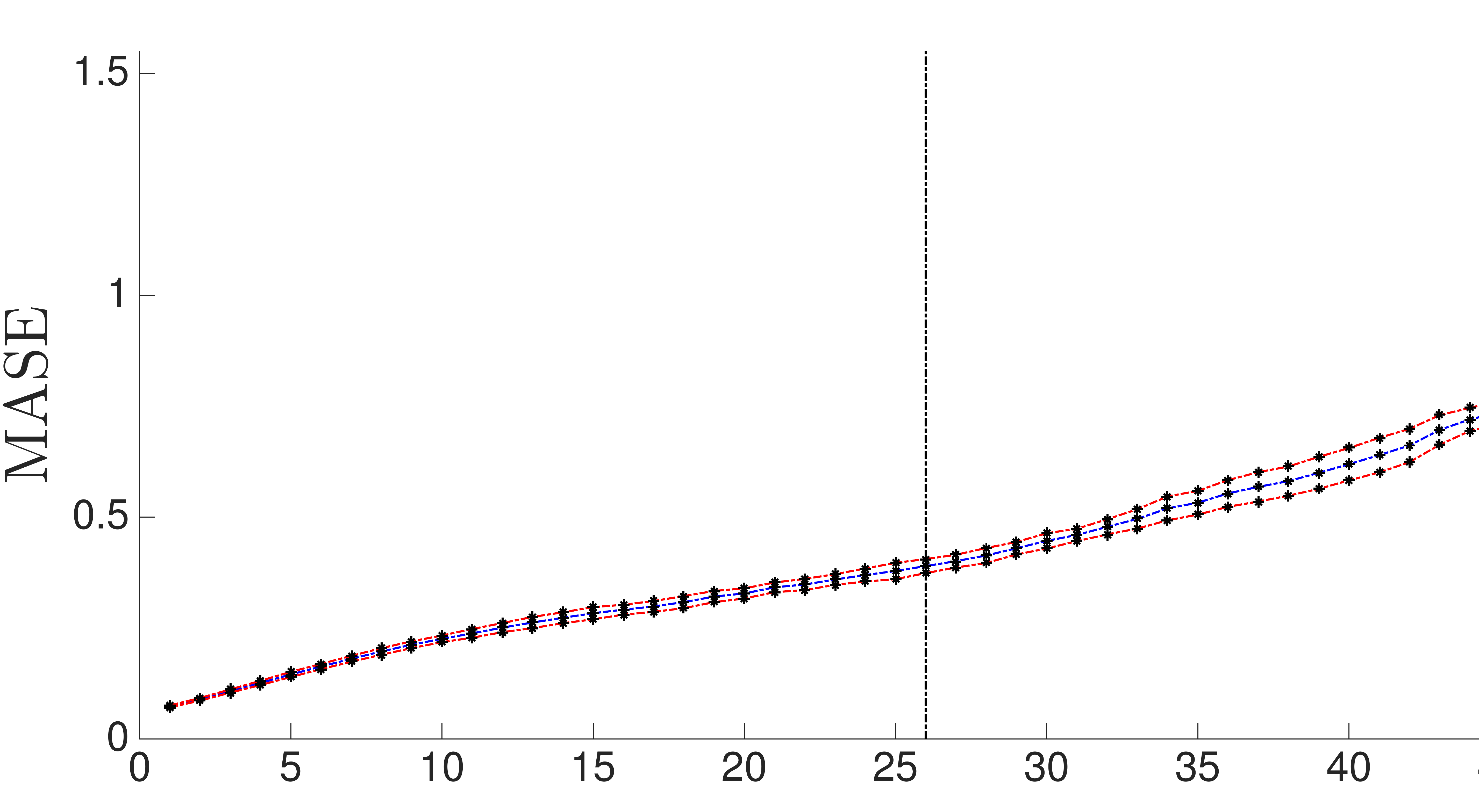}
                \caption{\roLMA on Lorenz-96 with $K=22$. }
        \end{subfigure}
        
\begin{subfigure}[b]{\columnwidth}
                \includegraphics[width=\columnwidth]{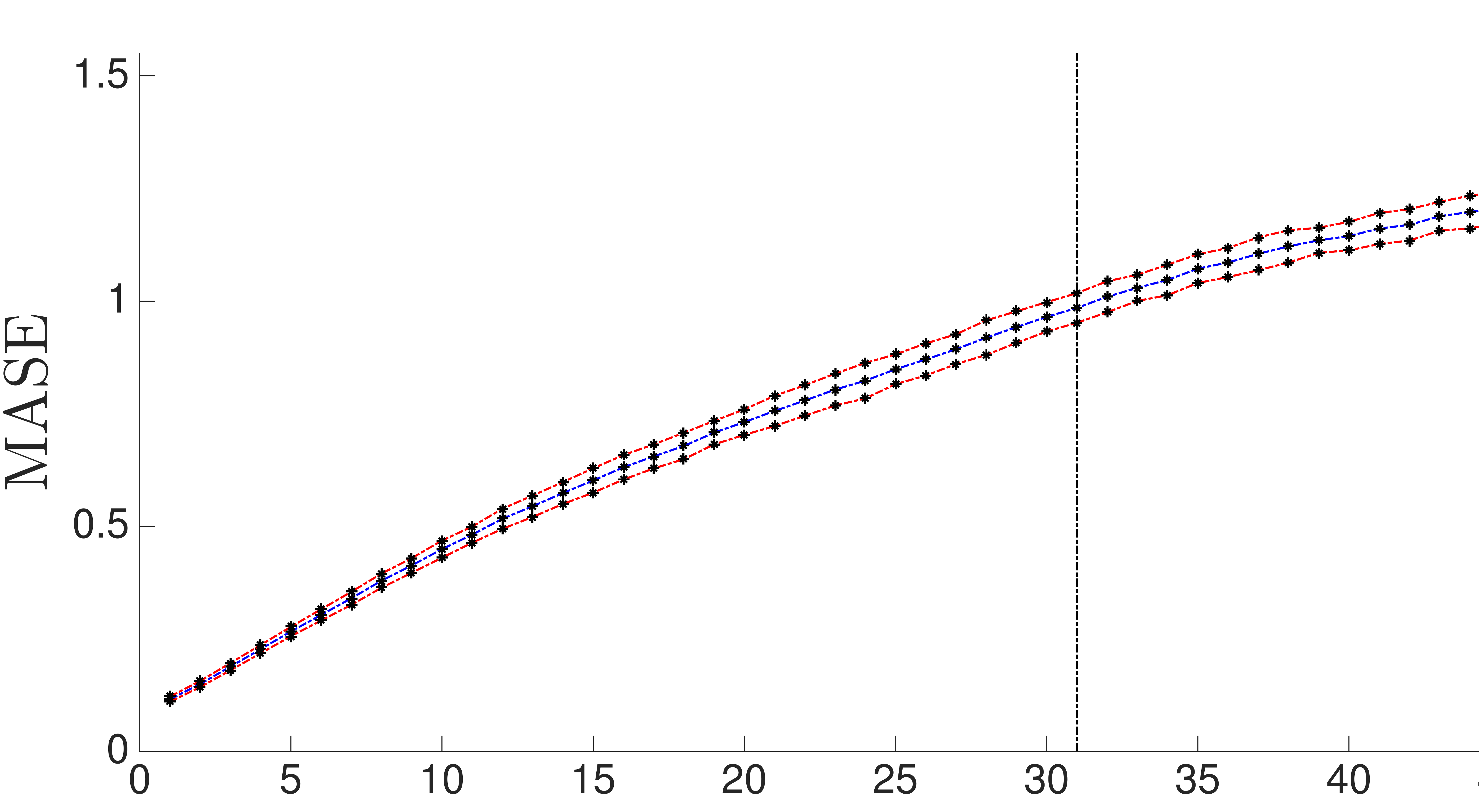}
                \caption{\roLMA on Lorenz-96 with $K=47$. }
        \end{subfigure}%
        
\begin{subfigure}[b]{\columnwidth}
                \includegraphics[width=\columnwidth]{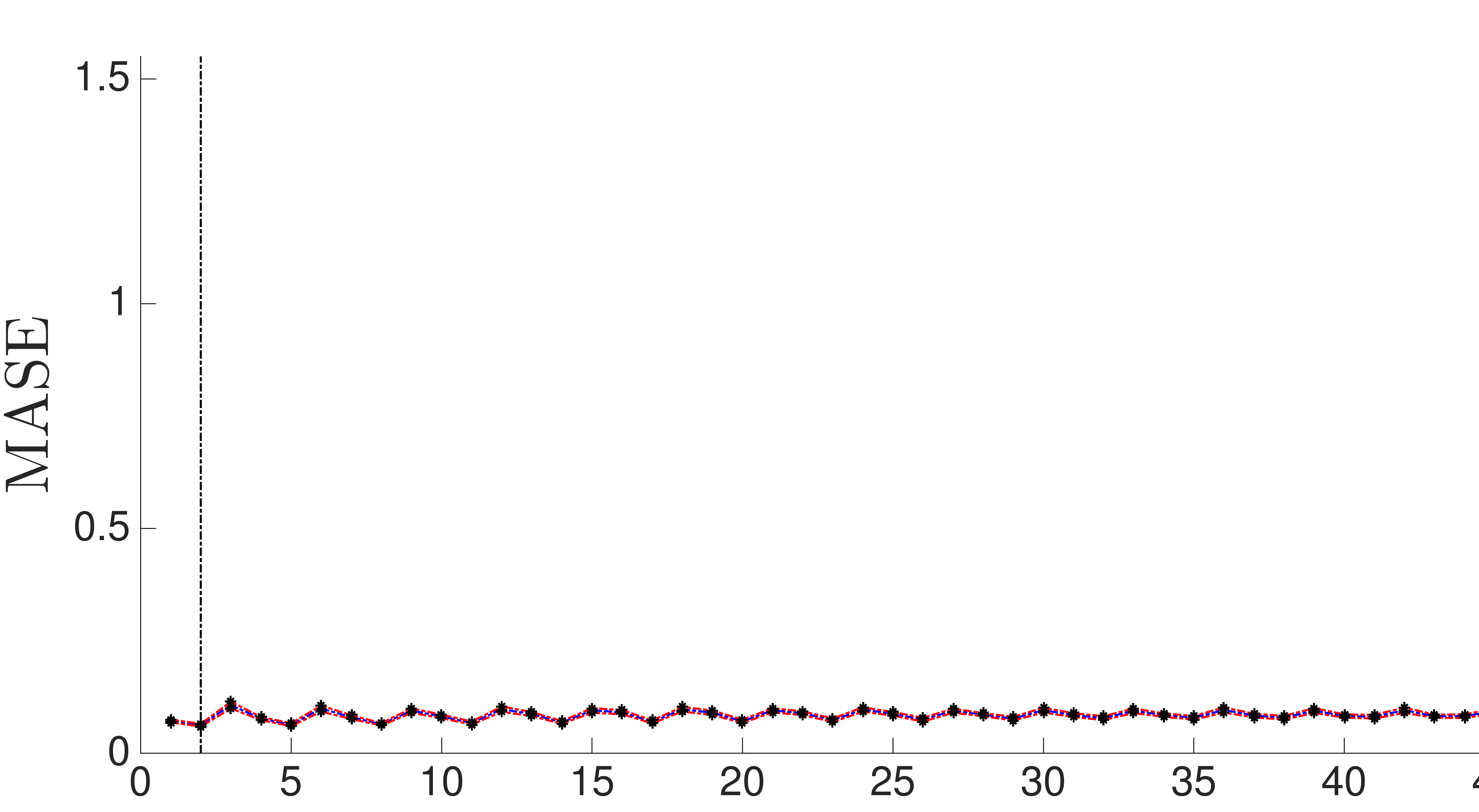}
                \caption{\roLMA on \col.}
        \end{subfigure}%
        
        \begin{subfigure}[b]{\columnwidth}
                \includegraphics[width=\columnwidth]{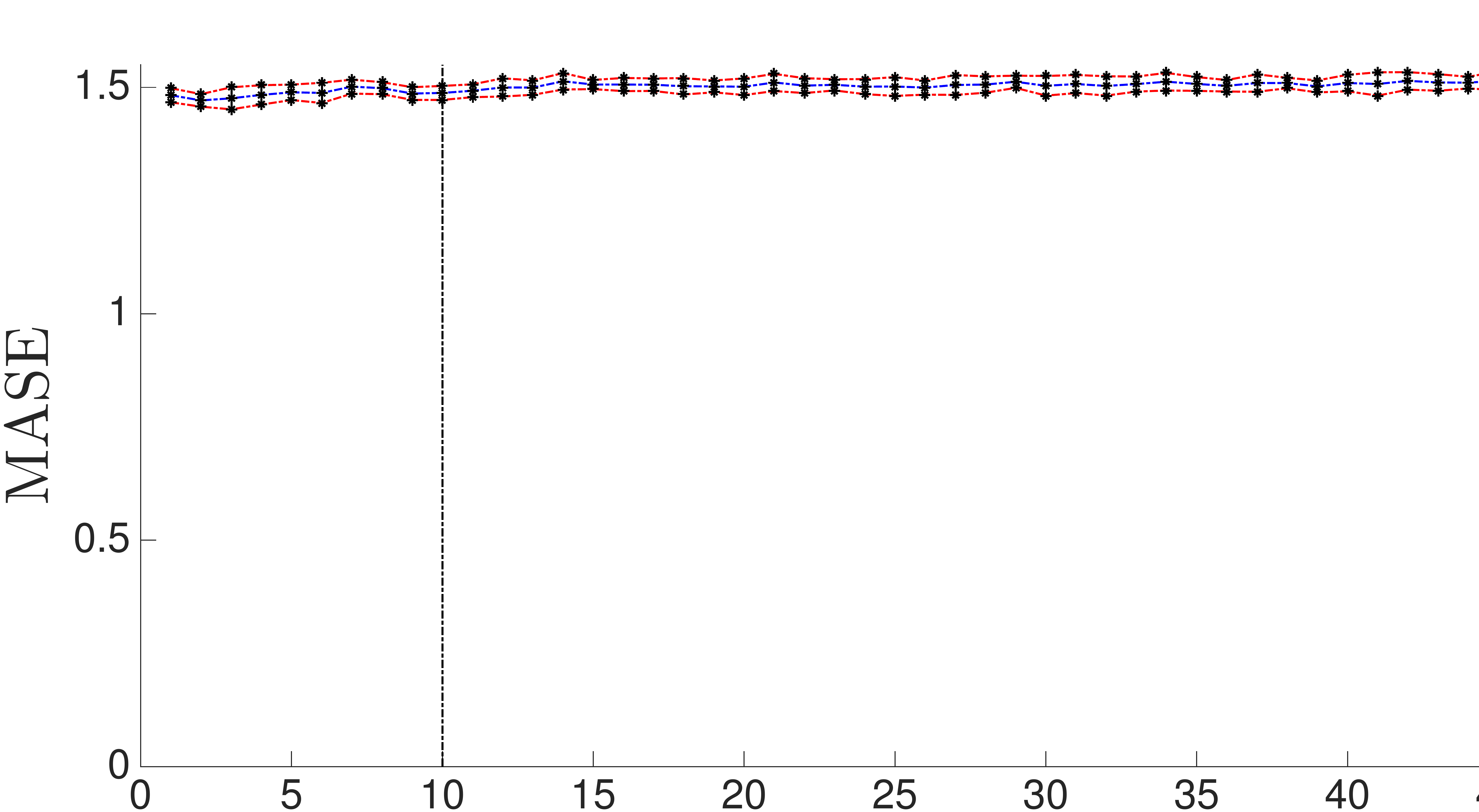}
                \caption{\roLMA on \gcc.}
        \end{subfigure}%
               
\caption{The effect of $\tau$ on \roLMA forecast accuracy.  The blue
dashed curves are the average $MASE$ of the \roLMA forecasts; the red
dotted lines show $\pm$ the standard deviation.  The black vertical
dashed lines mark the $\tau$ that is the first minimum of the mutual
information curve for all of the time
series.}
\label{fig:adapMASEvsTAU}
\end{figure}

Figure~\ref{fig:adapMASEvsTAU} explores this effect in more detail.
Across all $\tau$ values, the $MASE$ of \col was generally lower than
the other three experiments---again, simply because this time series
has more predictive structure.  The $K=22$ curve is generally lower
than the $K=47$ one for the same reason, as discussed at the end of
the previous section.  For the Lorenz-96 traces, prediction accuracy
decreases monotonically with $\tau$.  It is known that increasing
$\tau$ is beneficial for longer prediction horizons\cite{kantz97}.
Our situation involves short prediction horizons, so it makes sense
that our observation is consistent with the contrapositive of that
result.

For the experimental traces, the relationship between $\tau$ and
$MASE$ score is less simple.  There is only a slight upward overall
trend (not visible at this scale) and the curves are nonmonotonic.
This latter effect is likely due to periodicities in the dynamics,
which are very strong in the \col signal (viz., a dominant unstable
period-three orbit in the dynamics, which traces out the top, bottom,
and middle bands in Figure~\ref{fig:computerdata}).  Periodicities
cause obvious problems for embeddings---and forecast methods that
employ them---if the delay is a harmonic or subharmonic of those
periods, simply because the coordinates of the delay vector are not
independent samples of the dynamics.  It is for this reason that
Takens mentions this condition in his original paper.  Here, the
effect of this is an oscillation in the forecast accuracy vs. $\tau$
curve: low when it is a sub/harmonic of the dominant unstable periodic
orbit in the \col dynamics, for instance, then increasing with $\tau$
as more independence is introduced into the coordinates, then falling
again as $\tau$ reaches the next sub/harmonic, and so on.

This naturally leads to the issue of choosing a good value for the
delay parameter.  Recall that all of the experiments reported in
Section~\ref{sec:projresults} used a $\tau$ value chosen at the first
minimum of the mutual information curve for the corresponding time
series.  These values are indicated by the black vertical dashed lines
in Figure~\ref{fig:adapMASEvsTAU}.  This estimation strategy was
simply a starting point, chosen because it is arguably the most common
heuristic used in the nonlinear time-series analysis community.  As is
clear from Figure~\ref{fig:adapMASEvsTAU}, though, it is \emph{not}
the best way to choose $\tau$ for reduced-order forecast strategies.
Only in the case of \col is that $\tau$ value optimal
for \roLMA---that is, does it fall at the lowest point on the $MASE$
vs. $\tau$ curve.

\begin{figure}[ht!]
        \centering
        \begin{subfigure}[b]{0.95\columnwidth}
\includegraphics[width=0.95\columnwidth]{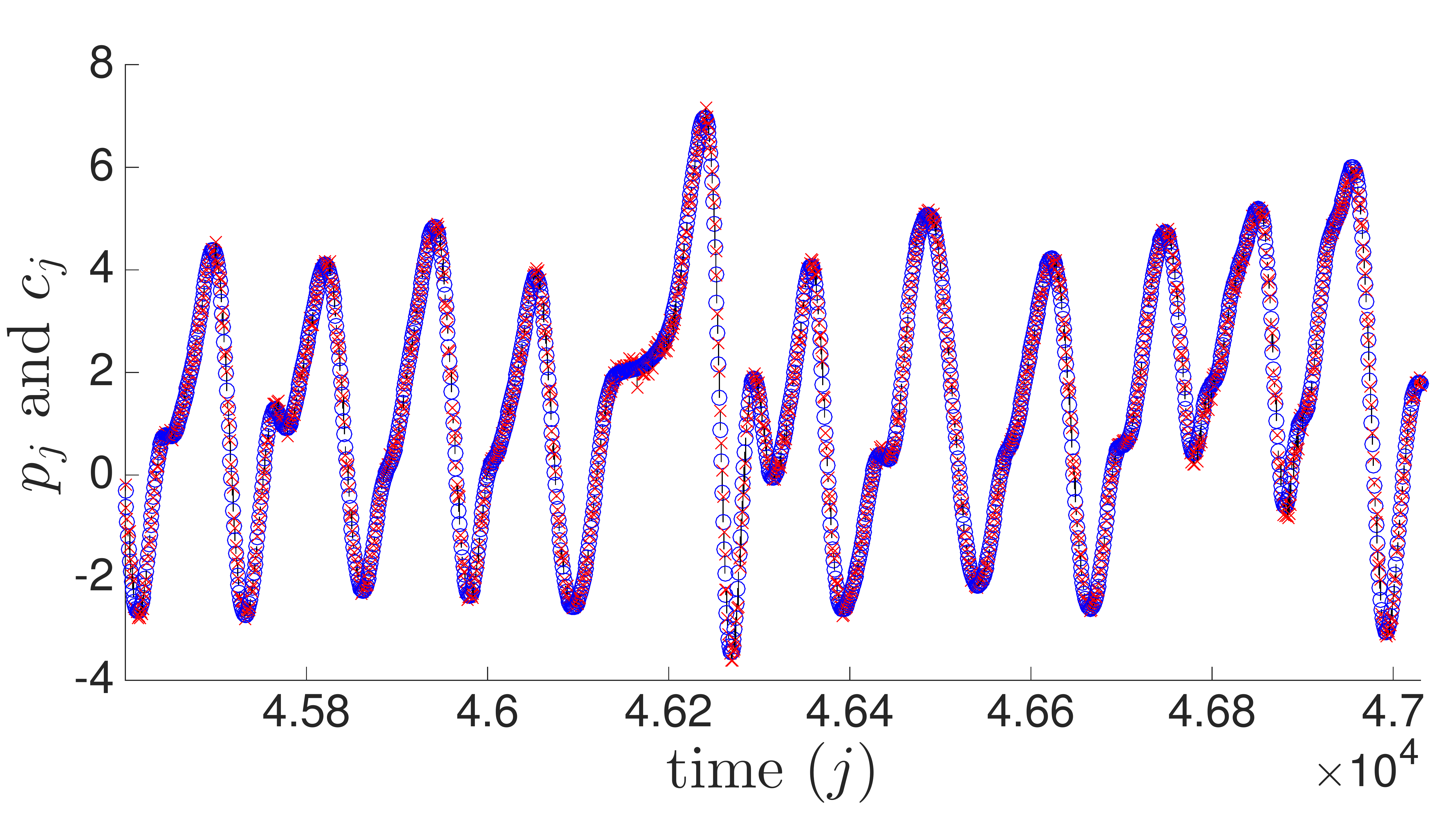}
                  \caption{A \roLMA forecast ($MASE=0.985$) using the ``default''
                  value of $\tau$ for this trace, which is chosen at the
                  first minimum of the average mutual information.}
        \end{subfigure}
        \begin{subfigure}[b]{0.95\columnwidth}
  \includegraphics[width=0.95\columnwidth]{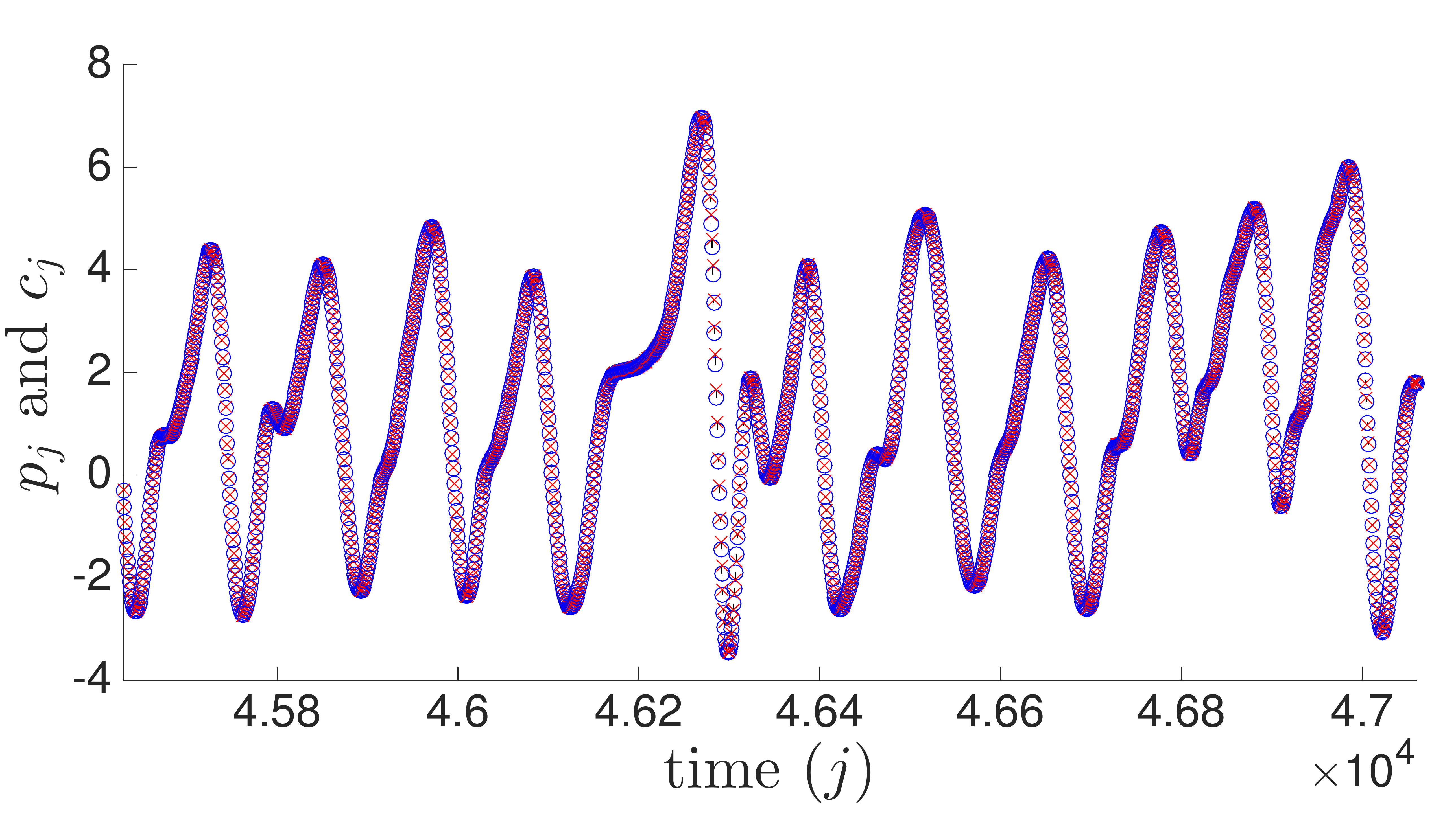}
                  \caption{A \roLMA forecast  ($MASE=0.115$) using the optimal
                  value of $\tau$ for this trace, which was chosen
                   at the
                  first minimum of the plot in Figure~\ref{fig:adapMASEvsTAU}(b).}
        \end{subfigure}
\caption{Time-domain plots of \roLMA forecasts of  a $K=47$ Lorenz-96
trace with default and best-case $\tau$ values.}
\label{fig:K47BestCasepredictions}
\end{figure}

This is the point alluded to at the end of
Section~\ref{sec:projresults}: one can improve the performance of
\roLMA  simply by choosing a different $\tau$---i.e., by adjusting the
one free parameter of the method.  In all cases (aside from \col,
where the default $\tau$ was the optimal value) adjusting $\tau$
brought \roLMA's error down below \fnnLMA's.  The improvement can be
quite striking: for visual comparison,
Figure~\ref{fig:K47BestCasepredictions} shows
\roLMA forecasts of a
$K=47$ Lorenz-96 trace using default and best-case values of $\tau$.
Again, this supports the main point of this paper: forecast methods
based on incomplete embeddings of time-series data can be very
effective---and much less work than those that require a full
embedding.
%
\begin{table*}
\caption{The effects of the $\tau$ parameter.  The ``default'' value 
is fixed, for both \roLMA and \fnnLMA, at the first minimum of the
average mutual information for that trace; the ``best case'' value is
chosen individually, for each method and each trace, from plots like
the ones in Figure~\ref{fig:adapMASEvsTAU}.}

  \begin{center}
  \begin{tabular}{lcccc}
 \hline Signal & \roLMA              & \roLMA               & \fnnLMA             & \fnnLMA\\
  & (default $\tau$)  & (best-case $\tau$) & (default $\tau$)  & (best-case $\tau$) \\ 
  \hline
    Lorenz-96 $K=22$     & $0.391\pm0.016$   & $0.073 \pm 0.002$  & $0.441 \pm 0.033$ & $0.137\pm0.006$ \\
    Lorenz-96 $K=47$     & $0.985 \pm 0.047$ & $0.115 \pm 0.006$  & $1.007 \pm 0.043$ & $0.325\pm0.020$ \\
    \col          & $0.063\pm0.003$   & $0.063\pm 0.003$& $ 0.050 \pm0.002$ & $0.049\pm0.002$ \\
    \gcc          & $1.488\pm0.016$   & $1.471\pm0.014$   & $ 1.530\pm 0.021$ &$1.239\pm0.020$\\ \hline
  \end{tabular}
  \end{center}
 \label{tab:best-case-error}
  \end{table*}

However, that comparison is not really fair.  Recall that the ``full''
embedding that is used by \fnnLMA, as defined so far, fixes $\tau$ at
the first minimum of the average mutual information for the
corresponding trace.  It may well be the case that that $\tau$ value
is suboptimal for \emph{that} method as well---as it was for \roLMA.
To make the comparison fair, we performed an additional set of
experiments to find the optimal $\tau$ for \fnnLMA.
Table~\ref{tab:best-case-error} shows the numerical values of the
$MASE$ scores, for forecasts made with default and best-case $\tau$ values, for both
methods and all traces.  In the two simulated examples,
best-case \roLMA ~ significantly outperformed best-case \fnnLMA; in
the two experimental examples, the best-case \fnnLMA was better, but
not by a huge margin.  That is, even if one optimizes $\tau$
individually for these two methods, \roLMA keeps up with, and sometimes
outperforms, \fnnLMA.

In view of our claim that part of \roLMA's advantage stems from the
natural noise mitigation effects of a low-dimensional embedding, it
may appear somewhat odd that \fnnLMA works better on the experimental
time-series data, which certainly contain noise.  Comparisons of large
$MASE$ scores are somewhat problematic, however.  
\label{page:suspect}
Recall that $MASE>1$ means that the forecast is worse than an
in-sample random-walk forecast of the same trace.  That is, both
LMA-based methods---no matter the $\tau$ values---generate poor
predictions.  There could be a number of reasons for this poor
performance.  \gcc has almost no predictive structure \cite{josh-pre}
and \fnnLMA's extra axes may add to its ability to capture that
structure---in a manner that outweighs the potential noise effects of
those extra axes.  The dynamics of \col, on the other hand, are fairly
low dimensional and dominated by a single unstable periodic orbit; it
could be that the ``full'' embedding of these dynamics captures its
structure so well that \fnnLMA is basically perfect and \roLMA cannot
do any better.


While the plots and $MASE$ scores in this paper
suggest that \roLMA forecasts are quite good, it is important to note
that both ``default'' and ``best-case'' $\tau$ values were chosen
after the fact in all of those experiments.  This is not useful in
practice.  A large part of the point of \roLMA is its ability to work
`on the fly,' when one may not have the leisure to run an average
mutual information calculation on a long segment of the trace and find
a clear minumum---and one \emph{certainly} cannot run a set of
experiments like the ones that produced Figure~\ref{fig:adapMASEvsTAU}
and choose an optimal $\tau$.  (Producing this Figure required 3,000
runs involving a total of 22,010,700 forecasted points, which took
approximately 44.5 hours on an Intel Core i7.)

We suspect, though, that it is possible to estimate good---although
certainly not optimal---values for $\tau$, and to do so quickly and
automatically.  Most of the current $\tau$-estimation methods use some
kind of distribution.  We are exploring how to adapt them to be used
in a ``rolling'' fashion, on an evolving distribution that is built up
on the fly, as the data arrive.  Newer variations on mutual
information may be more appropriate in this situation, such as
co-information\cite{Bell03theco-information} and
multi-information\cite{Studeny98themultiinformation}.  An appealing
alternative is a time-lagged version of permutation
entropy\cite{PhysRevE.70.046217}.  We are also exploring geometric and
topological heuristics from the nonlinear time-series analysis
literature, such as \emph{wavering
product} \cite{liebert-wavering}, \emph{fill factor}, \emph{integral
local deformation}\cite{Buzugfilldeform}, and \emph{displacement from
diagonal}\cite{rosenstein94}.  Many of these methods, however, are
aimed at producing reconstructions from which one can accurately
estimate dynamical invariants; as mentioned above, the optimal $\tau$
for \emph{forecasting} may be quite different.  Moreover, any method that
works with the geometry or topology of the dynamics---rather than a
distribution of scalar values---may be less able to work with short
samples of that structure.  Nonetheless, these methods may be a useful
complement to the statistical methods mentioned above.

\subsubsection{Prediction horizon}

There are fundamental limits on the prediction of chaotic systems.
Positive Lyapunov exponents make long-term forecasts a difficult
prospect beyond a certain point for even the most-sophisticated
methods\cite{weigend-book, kantz97, josh-pre}.  Note that the
coordinates of points in higher-dimensional embedding spaces sample
wider temporal spans of the time series.  This increased `memory'
means that they capture more information about the dynamics than
lower-dimensional embeddings do, which raises an important concern
about \roLMA: whether its accuracy will degrade more rapidly with
increasing prediction horizon than that of \fnnLMA.

Recall that the formulations of both methods, as described and
deployed in the previous sections, assume that measurements of the
target system are available in real time: they ``rebuild'' the LMA
models after each step, adding new time-series points to the
embeddings as they arrive.  Both \roLMA and \fnnLMA can easily be
modified to produce longer forecasts, however---say, $h$ steps at a
time, only updating the model with new observations at $h$-step
intervals.  Naturally, one would expect forecast accuracy to suffer as
$h$ increases for any non-constant signal.  The question at issue in
this section is whether the greater temporal span of the data points
used by \fnnLMA mitigates that degradation, and to what extent.

\begin{table*}
\caption{The $h$-step mean
absolute scaled error ($h$-$MASE$) scores for different forecast
horizons ($h$).  As explained in the text, $h$-$MASE$ scores should
not be compared for different $h$ (i.e., down the columns of this
table).}
  \begin{center}
  \begin{tabular}{ccccccc}
  \hline\hline Signal  & $h$ & \roLMA & \roLMA & \fnnLMA & \fnnLMA \\ 
& & (default $\tau$) & (best-case $\tau$) & (default $\tau$) &  (best-case $\tau$) \\
  \hline
Lorenz-96 $K=22$  &1& $0.391\pm0.016$ & $0.073\pm0.002$  & $0.441\pm0.003$ & $ 0.137\pm0.006$ \\
Lorenz-96 $K=22$ &10& $0.101\pm0.008$ & $0.066\pm0.003$   & $0.062\pm0.011$ & $0.033\pm0.002$ \\
Lorenz-96 $K=22$ &50& $0.084\pm0.007$ & $0.074\pm0.008$  &  $0.005\pm0.002$ & $0.004\pm0.001$ \\
Lorenz-96 $K=22$ &100& $0.057\pm0.005$& $0.050\pm0.004$ & $0.003\pm0.001$ & $0.003\pm0.001$ \\
\hline
Lorenz-96 $K=47$  &1& $0.985\pm0.047$ & $0.115\pm0.006$ & $0.995\pm0.053$ & $0.325\pm 0.020 $ \\
Lorenz-96 $K=47$  &10 &$0.223\pm0.011$  &  $0.116\pm 0.005$  & $ 0.488\pm0.042$& $0.218\pm0.012$ \\
Lorenz-96 $K=47$  &50 &$0.117\pm0.011$&$0.112\pm 0.010$& $0.127\pm0.011$& $0.119\pm0.010$\\

Lorenz-96 $K=47$  &100 &$0.075\pm0.006$ &$0.068\pm0.005$ & $0.079\pm0.005$& $0.075\pm0.004$ \\

\hline
\col  &1 &$0.063\pm 0.003$&$0.063\pm 0.003$ &$0.050\pm0.002$ & $0.049\pm0.002$  \\
\col  &10 & $0.054\pm0.006$& $0.046\pm0.003$& $0.021\pm0.001$& $0.018\pm0.001$ \\
\col  &50 & $0.059\pm0.009$& $0.037\pm0.003$& $0.012\pm0.003$& $0.009\pm0.001$ \\
\col  &100 & $ 0.044\pm0.004$& $0.028\pm0.006$& $0.010\pm0.003$& $0.007\pm0.001$ \\
\hline
\gcc  &1 &$1.488\pm0.016$&$1.471\pm0.014$&$1.530\pm0.021$&$1.239\pm0.020$\\
\gcc  &10 & $0.403\pm0.009$& $0.396\pm0.009$ & $0.384\pm0.007$ & $0.369\pm0.010$   \\
\gcc  &50 &$0.154\pm0.003$& $0.151\pm0.005$& $0.143\pm0.003$ & $0.141\pm0.003$ \\
\gcc  &100 & $0.101\pm0.002$& $0.101\pm0.003$  &$0.095\pm0.002$& $0.093\pm0.002$  \\
  \hline\hline
  \end{tabular}
  \end{center}
 \label{tab:hstep}
  \end{table*}

Assessing the accuracy of $h$-step forecasts requires a minor
modification to the $MASE$ calculation, since its denominator is
normalized by \emph{one-step} random-walk forecast errors.  In an
$h$-step situation, one should instead normalize by $h$-step
forecasts, which involves modifying the scaling term to be the average
root-mean-squared error accumulated using random-walk forecasting on
the initial training signal, $h$ steps at a time\cite{MASE}.  This
gives a new figure of merit that we will call $h$-$MASE$:
$$h\text{-}MASE = \sum_{j=n+1}^{k+n+1}\frac{|p_j-c_j|
}{\frac{k}{n-h}\sum^n_{i=1}\sqrt{\frac{\sum^h_{\iota=1}(x_{i}-x_{i+\iota})^2}{h}}}$$
Note that the $h$-step forecast accuracy of the random-walk method
will also degrade with $h$, so $h$-$MASE$ will always be lower than
$MASE$.  This also means that $h$-$MASE$ scores should not be compared
for different $h$.

In Table~\ref{tab:hstep}, we provide $h$-$MASE$ scores for $h$-step
forecasts of the different sets of experiments from
Section~\ref{sec:projresults}.  The important comparisons here are, as
mentioned above, across the rows of the table.  The different methods
``reach'' different distances back into the time series to build the
models that produce those forecasts, of course, depending on their
delay and dimension.  At first glance, this might appear to make it
hard to compare, say, default-$\tau$ \roLMA and
best-case-$\tau$ \fnnLMA sensibly, since they use different $\tau$s
and different values of the embedding dimension and thus are spanning
a longer range of the time series.  However, $h$ is measured in units
of the sample interval of the time series, so comparing one $h$-step
forecast to another (for the same $h$) does make sense.

There are a number of interesting questions to ask about the patterns
in this table, beginning with the one that set off these experiments:
how do \fnnLMA and \roLMA compare if one individually optimizes $\tau$
for each method?  The numbers indicate that \roLMA beats \fnnLMA for
$h=1$ on the $K=22$ traces, but then loses progressively badly (i.e.,
by more $\sigma$s) as $h$ grows.  \col follows the same pattern except
that \roLMA is worse even at $h=1$.  For \gcc, \fnnLMA performs better
at both $\tau$s and all values of $h$, but the disparity between the
accuracy of the two methods does not systematically worsen with
increasing $h$.  For $K=47$, \roLMA consistently beats \fnnLMA for
both $\tau$s for $h\leq 10$ but the accuracy of the two methods is
comparable for longer prediction horizons.

Another interesting question is whether the assertions in the previous
section stand up to increasing prediction horizon.  Those assertions
were based on the results that appear in the $h=1$ rows of
Table~\ref{tab:hstep}: \roLMA was better than \fnnLMA on the $K=22$
Lorenz-96 experiments, for instance, for both $\tau$ values.  This
pattern does not persist for longer prediction horizons:
rather, \fnnLMA generally outperforms \roLMA on the $K=22$ traces for
$h=10, 50,$ and 100.  The $h=1$ comparisons for $K=47$
and \col \emph{do} generally persist for higher $h$, however.  As
mentioned before, \gcc is problematic because its $MASE$ scores are so
high, but the accuracies of the two methods are similar for all $h>1$.

\begin{figure*}[ht!]
        \centering
\includegraphics[width=1.8\columnwidth]{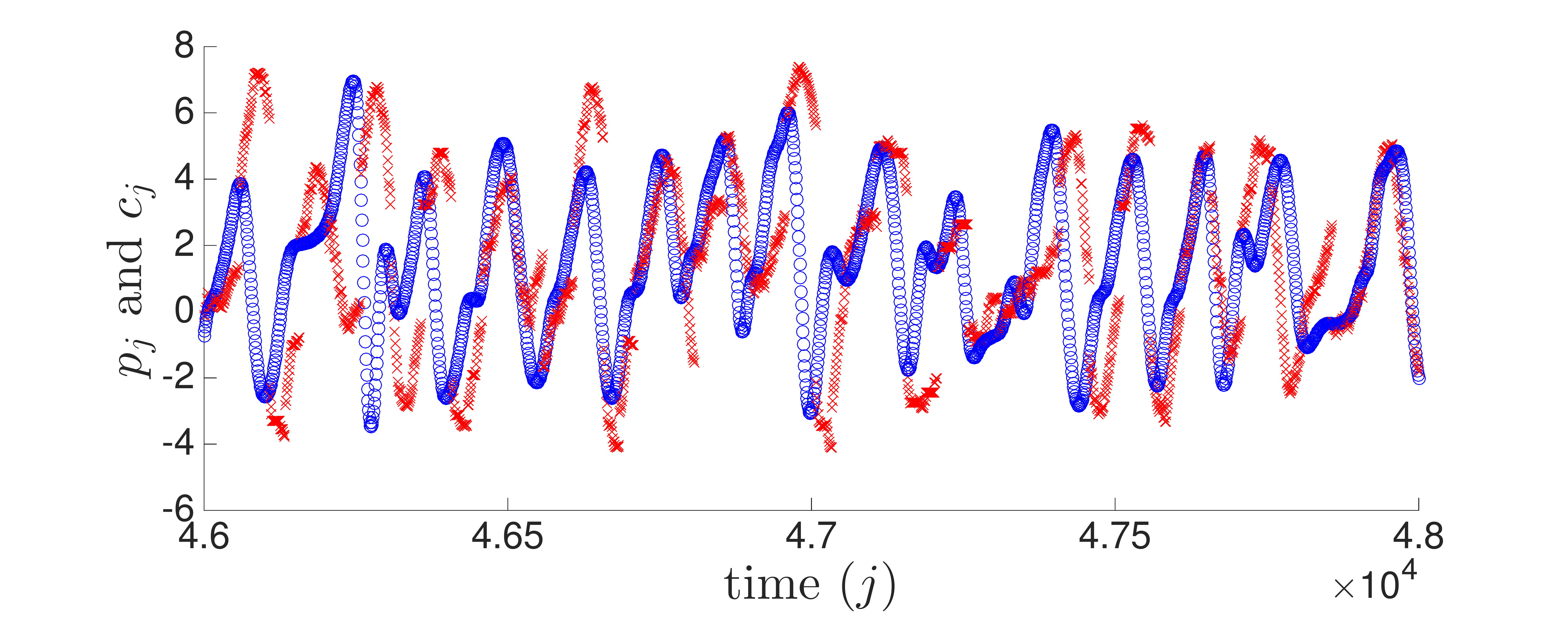}

\caption{A  best-case-$\tau$ \roLMA forecast of a $K=47$ Lorenz-96 
trace for $h=50$.  The forecast (red) follows the true trajectory
(blue) for a while, then falls off onto a shadow trajectory, then gets
recorrected when a new set of observations are incorporated into the
model after $h$ time steps.}
\label{fig:shadowtrajectories}
\end{figure*}

The fact that \fnnLMA generally outperforms \roLMA for longer
prediction horizons makes sense, simply because \roLMA samples less of
the time series and therefore has less `memory' about the dynamics.
Still, best-case \roLMA performs almost as well as best-case \fnnLMA
in many cases, even for $h=100$.  In view of the fundamental limits on
prediction of chaotic dynamics, however, it is worth considering
whether \emph{either} method is really making correct long-term
forecasts.  Indeed, time-domain plots of long-term forecasts (e.g.,
Figure~\ref{fig:shadowtrajectories}) reveal that both \fnnLMA
and \roLMA forecasts have fallen off the true trajectory and onto
shadow trajectories---a well-known phenomenon when forecasting chaotic
dynamics\cite{sauer-delay}. 

In other words, it appears that even a 50-step forecast of these
chaotic trajectories is a tall order: i.e., that we are running up
against the fundamental bounds imposed by the Lyapunov exponents.  In
view of this, it is promising that \roLMA generally keeps up
with \fnnLMA in many cases---even when both methods are struggling
with the prediction horizon, and even though the \roLMA model has much
less memory about the past history of the trajectory.  An important
aspect of our future research on this topic will be determining bounds
on reasonable prediction horizons---as well as developing methods, if
possible, to increase prediction horizon without sacrificing the
accuracy or speed of \roLMA.

\section{Conclusion}\label{sec:concl}

We have proposed a novel nonlinear forecast strategy that works in a
two-dimensional version of the delay-coordinate embedding space.  Our
preliminary results suggest that this approach captures the dynamics
well enough to enable effective prediction of several very different
real-world and artificial dynamical systems, even though working with
a $2D$ embedding violates one of the most critical basic tenets of the
delay-coordinate embedding machinery.

The point of this paper is not only to explore whether prediction in
projection works, but also to establish it as a useful practical
technique.  From that standpoint, the primary advantage of \roLMA is
that the $2D$ embedding that it uses to model the dynamics has only a
single free parameter: the delay, $\tau$.  The standard
delay-coordinate embedding process has a second free parameter (the
dimension) that requires expert human judgment to estimate, making
that class of methods all but useless for adaptive modeling and
forecasting.  As described at the end of
Section~\ref{sec:varyingproj}, we believe that good values for $\tau$
can be effectively estimated on the fly from the time series.  This
will let our method adapt to nonstationary dynamics.  In this fashion,
the line of research described in this paper bridges the gap between
rigorous nonlinear mathematical models---which are ineffective in
real-time---and approximate methods that are agile enough for adaptive
modeling of nonstationary dynamical processes.

Data length is an important consideration in any nonlinear time-series
application.  Traditional estimates (\emph{e.g.,} by
Smith\cite{smithdatabound} and by Tsonis \emph{et
al.}\cite{tsonisdatabound}) would suggest that $\approx 10^{17}$ data
points would be required for a successful delay-coordinate embedding
for the Lorenz-96 $K=47$ data in Section~\ref{sec:roLMALorenz96},
where the known $d_{KY}$ values\cite{KarimiL96} indicate that one
might need at least $m=38$ dimensions to properly unfold the dynamics.
While it may be possible to collect that much data in a synthetic
experiment, that is certainly not an option in a real-world
forecasting situation, particularly if the dynamics that one wants to
forecast are nonstationary.  Note, however, that we were able to get
good results on that system with only 45,000 points.  The Smith/Tsonis
estimates were derived for the specific purposes of correlation
dimension calculations via the Grassberger-Procaccia algorithm; in our
opinion, they are overly pessimistic for forecasting.  For example,
Sauer~\cite{sauer-delay} successfully forecasted the continuation of a
16,000-point time series
embedded in 16 dimensions; Sugihara \& May~\cite{sugihara90} used
delay-coordinate embedding with $m$ as large as seven to successfully
forecast biological and epidemiological time-series data as short as
266 points.
%
%
Given these results, we believe that our traces are long enough to
support the conclusions that we drew from them: viz., \fnnLMA is a
reasonable point of comparison, and the fact that \roLMA outperforms
it is meaningful.  Nonetheless, an important future-work item will be
a careful study of the effects of data length on \roLMA, which will
have profound implications for its ability to handle nonstationarity.

As stated before, no forecast model is ideal for all noise-free
deterministic signals, let alone all real-world time-series data sets.
However, the proof of concept offered in this paper is encouraging:
prediction in projection appears to work remarkably well, even though
the models that it uses are not topologically faithful to the true
dynamics.  As mentioned in the Introduction, there are many other
creative and effective ways to leverage the structure of an embedded
dynamics in order to predict the future course of a trajectory ({\it
e.g.},~\cite{casdagli-eubank92,weigend-book,Smith199250,sauer-delay,sugihara90,pikovsky86-sov}).
It would be interesting to see how well these methods work in a
reduced-order embedding space.  Our ultimate goal is to be able to
show that prediction in projection---a simple yet powerful reduction
of a time-tested method---has real practical utility for a wide
spectrum of forecasting tasks as a simple, agile, adaptive,
noise-resilient, forecasting strategy for nonlinear systems.

\bibliographystyle{unsrt}

 \bibliography{master-refs}
\end{document}